\documentclass[prb,twocolumn,showpacs,preprintnumbers,amsmath,amssymb,superscriptaddress]
{revtex4}
\usepackage{soul}
        \setul{0.5ex}{}
\usepackage{graphicx}
\usepackage[center]{subfigure}
\usepackage{color}
\usepackage[usenames,dvipsnames,svgnames,table]{xcolor}
\usepackage{rotating}
\usepackage{placeins}
\usepackage{verbatim}
\usepackage{float}
\usepackage{ulem}
\newcommand \be {\begin{equation}}
\newcommand \ee {\end{equation}}
\newcommand \ben {\begin{eqnarray}}
\newcommand \een {\end{eqnarray}}

\newcommand \liqr {\rho_{\ell}}

\begin{document}

\title{
An Atomistic Study of 
Diffusion-Mediated Plasticity and Creep\\
using Phase Field Crystal Methods
}

\author{Joel Berry}
\altaffiliation[]{{\it Current address}:
Department of Mechanical and Aerospace Engineering,
Princeton University, Princeton, NJ 08544, USA}
\affiliation{Department of Materials Science and Engineering,
McMaster University, 1280 Main Street West, Hamilton, Ontario, L8S 4L7, Canada}
\affiliation{
Department of Physics and Astronomy, The University of British Columbia, 
6224 Agricultural Road, Vancouver, British Columbia, V6T 1Z1, Canada}
\author{J\"org Rottler}
\affiliation{
Department of Physics and Astronomy, The University of British Columbia, 
6224 Agricultural Road, Vancouver, British Columbia, V6T 1Z1, Canada}
\author{Chad W.\ Sinclair}
\affiliation{
Department of Materials Engineering, The University of British Columbia, 
309-6350 Stores Road, Vancouver, British Columbia, 
V6T 1Z4, Canada}
\author{Nikolas Provatas}
\affiliation{Physics Department, McGill University,
3600 rue University, Montr\'eal, Qu\'ebec, H3A 2T8, Canada}

\date{\today}

\begin{abstract}
The nonequilibrium dynamics of 
diffusion-mediated plasticity and creep
in materials subjected to constant load at high homologous temperatures 
is studied atomistically
using Phase Field Crystal (PFC) methods.
Creep stress and grain size exponents
obtained for nanopolycrystalline systems, 
$m \simeq 1.02$ and $p \simeq 1.98$, respectively,
closely match those expected for idealized diffusional Nabarro-Herring creep.
These exponents are observed in the presence of
significant stress-assisted diffusive grain boundary migration,
indicating that Nabarro-Herring creep and 
stress-assisted boundary migration contribute in the same manner
to the macroscopic constitutive relation.
When plastic response is dislocation-mediated,
power law stress exponents inferred from dislocation climb rates are found to
increase monotonically from $m \simeq 3$, 
as expected for generic climb-mediated natural creep,
to $m \simeq 5.8$ as the dislocation density
$\rho_d$ is increased beyond typical experimental values.
Stress exponents $m \gtrsim 3$ directly measured from simulations that
include dislocation nucleation, climb, glide, and annihilation
are attributed primarily to these large $\rho_d$ effects.
Extrapolation to lower $\rho_d$ suggests
that $m \simeq 4-4.5$ should be obtained from our PFC description
at typical experimental $\rho_d$ values, which is consistent with 
expectations for power law creep via mixed climb and glide.
The anomalously large stress exponents observed in our atomistic simulations
at large $\rho_d$ may nonetheless be relevant
to systems in which comparable densities are obtained locally
within heterogeneous defect domains such as dislocation cell walls or tangles.
\end{abstract}

\pacs{
61.72.Bb, 
61.72.Lk, 
62.20.F- 
81.40.Lm  
}
\maketitle

The tendency of a solid material to gradually and irreversibly deform
or even flow
under low loads
at high temperatures
is termed creep deformation 
\cite{kassnercreepbook,defmapsbook82,nabarrocreepbook95,
rajashby71,herring1950,coble1963}.
Low loads and high temperatures in this context are 
$\sigma \lesssim \sigma_y$ and
$T \gtrsim T_m/2$, respectively,
where $\sigma_y$ is the yield stress
and $T_m$ is the equilibrium melting temperature of the material.
This type of slow plasticity
not only alters material microstructure, shape, and properties over 
extended time periods, but is also
a primary cause of mechanical failure in materials
such as gas turbine blades 
that operate in high temperature load bearing environments
\cite{kassnercreepbook,defmapsbook82,nabarrocreepbook95}.

The plastic flow that occurs during creep 
can involve conservative defect 
evolution
mechanisms, 
such as dislocation glide and grain boundary sliding,
but is generally facilitated by thermally
activated defect motion along local stress and/or chemical potential
gradients, and is therefore
inherently diffusive in nature.
Vacancy diffusion in particular tends to be a central facilitator
of deformation, either directly during diffusional creep
or indirectly during dislocation or power law creep,
as discussed further in the following.
Moreover, it is the collective evolution of different defect populations
over multiple length and time scales that leads to the observed
diverse macroscopic phenomenology of creep.
If the characteristic time scales associated with the diffusion of
individual defects
are almost entirely inaccessible to most atomistic descriptions, then
the length and time scales associated with
collective defect diffusion and macroscopic
deformation are decidedly unreachable with conventional approaches.
Details of the atomic-scale mechanisms by which 
diffusional (low $\sigma$)
and, in particular, power law (higher $\sigma$) creep occur
therefore remain matters of speculation 
in many cases.
There is clearly a need to better characterize
these mechanisms,
to understand how they collectively generate
mesoscale dislocation pattering phenomena observed during creep, 
and to understand how these mesoscale structures connect to macroscopic
creep phenomenology.

The aims of the first part of the present study are to 
reproduce the macroscopic phenomenology of
diffusional creep 
in nanopolycrystals from long time scale atomistic simulations,
and to examine the consequences of 
stress-assisted grain boundary migration
\cite{kimSAGG99,kimSAGG00,haslamSAGG03,CahnTaylorGBcoupling04,GBmotionMDPFC12}
on diffusional creep.
The primary aim of the second part of this study 
is to begin addressing the question of whether such simulations
can also provide direct connections between 
atomistic defect evolution mechanisms and dislocation-mediated power law
creep phenomenology. 
Toward these ends we employ the phase field crystal (PFC) modeling approach.
We begin with a brief overview of creep phenomenology
and a discussion of simulation literature
in Section \ref{sec:creep}.
In Section \ref{sec:pfc}, the PFC approach is introduced and a new
method for conducting constant stress PFC creep simulations is outlined.
Grain boundary mediated deformation in dislocation-free nanopolycrystals
is then examined in Section \ref{sec:hexcreep}, 
while dislocation-mediated creep is examined
in Section \ref{sec:disloccreep}.
Connections to power law creep phenomenology are discussed
in Section \ref{sec:disloccreep},
and conclusions are summarized in Section \ref{sec:conclusions}.

\section{Creep Phenomenology}
\label{sec:creep}

The macroscopic plastic response of a material undergoing 
creep deformation can generally be characterized 
by an equation of the form
\cite{kassnercreepbook,defmapsbook82,nabarrocreepbook95}
\be
\frac{d\epsilon}{dt} \sim \left(\frac{\sigma}{\mu}\right)^m
\left(\frac{1}{d}\right)^p 
\label{creepeq}
\ee
where $\epsilon$ is total plastic strain, $t$ is time,
$\mu$ is shear modulus, $d$ is grain size,
and $m$ and $p$ are stress and grain size exponents, respectively.
In laboratory experiments, a constant stress is typically applied
and the resulting macroscopic deformation is measured as a function of time $t$.
The total strain can be decomposed into elastic and plastic contributions,
$\epsilon(t) = \epsilon^{\rm el} + \epsilon^{\rm pl}(t)$.
A rapid elastic deformation is thus observed, followed by an initial regime of
non-steady plastic flow called primary creep.
Plastic flow eventually reaches a steady-state during the so-called
secondary creep regime, such that the strain rate is constant in time
and Eq.\ (\ref{creepeq}) is applicable.

Diffusional creep, characterized by a stress exponent $m \simeq 1$,
is often observed at low $\sigma$ in such experiments.
At sufficiently high $T$, bulk vacancy diffusion is the dominant
flow mechanism and a grain size exponent $p \simeq 2$ results
(Nabarro-Herring creep \cite{herring1950}). 
At lower $T$, vacancy diffusion along grain boundaries is the dominant
flow mechanism and a grain size exponent $p \simeq 3$ results 
(Coble creep \cite{coble1963}).
Previous atomistic studies have addressed aspects of diffusional creep.
Coble creep
has been reproduced in molecular dynamics (MD) simulations 
of bcc \cite{creepYamakov2002,creepMillett2008} 
and fcc \cite{creepfccMD11} nanopolycrystals, for example.
This is feasible since vacancy diffusion rates can be relatively rapid
within loosely packed high angle grain boundaries.
Signs of non-negligible Nabarro-Herring creep
contributions have been reported as the grain size is increased 
\cite{creepMillett2008}, but
a clear transition to dominant Nabarro-Herring creep has not been
observed in MD simulations to our knowledge.

Deviations from the idealized descriptions of Nabarro-Herring and Coble 
creep are known to occur in systems with non-trivial microstructures.
For example, when all grains in a polycrystal have the same shape and size,
the idealized models of diffusional creep predict
a symmetric and affine elongation of all grains.
However, the presence of a distribution of grain sizes has been shown to 
lead inevitably to stress-assisted grain growth,
the shrinkage of some grains and growth of others, during
diffusional creep \cite{kimSAGG99}.
Thus, when grain boundaries are sufficiently mobile, idealized
diffusional creep behavior can be significantly modified by the simultaneous
effects of stress-assisted grain boundary migration.
Consequences include non-trivial, non-affine evolution of grain morphology
and the suppression of grain elongation for sufficiently large
grain boundary mobilities and low stresses \cite{kimSAGG00}.
These effects have been examined with both MD \cite{haslamSAGG03}
and mesoscale simulation methods \cite{kimSAGG00}.

At applied stresses higher than those at which diffusional creep is
typically observed, 
significant dislocation motion
begins to occur as vacancy fluxes around dislocation cores and
rates of stress-assisted thermally activated motion increase.
A transition from diffusional to 
$m>1$ or
power law creep then occurs, 
such that 
$m \simeq 3-10$,
with the precise value 
depending on a variety of material and microstructural properties,
many of which are not well understood.
Simultaneous interactions involving vacancies, dislocations, 
grain boundaries, and other defects
can lead to complex behaviors such as climb-assisted obstacle bypass
\cite{climblevelset10}
on atomistic scales and
collective dislocation patterning 
\cite{collectivedislocs2007}
on mesoscopic scales.
All manner of such processes may contribute to macroscopic 
power law creep rates. 

The simplest description of dislocation-mediated power law creep, 
$m=3$ natural creep \cite{nabarrocreepbook95},
is arrived at by assuming a linear relation between the steady-state 
dislocation velocity and stress, $v_{\rm ss} \sim \sigma$,
and by also employing a phenomenological relation between 
dislocation density $\rho_d$ 
and stress (Taylor's relation), $\rho_d \sim \sigma^2$.
Orowan's equation for plastic strain then specifies
$\epsilon^{\rm pl}(t) = 
\frac{1}{V} \sum^N_{i=1} ( b_i \ell_i \delta x_i )$,
where $V$ is the system volume, $N$ is the number of mobile dislocations,
$b_i$ is the magnitude of the Burgers vector of dislocation $i$,
$\ell_i$ is its length,
and $\delta x_i$ is the distance that it has traversed at time $t$.
The corresponding incremental plastic strain rate is
$\dot{\epsilon}^{\rm pl}(t) = 
\frac{1}{V} \sum^N_{i=1} ( b_i \ell_i v_i )$,
where $v_i$ is the velocity of dislocation $i$ at time $t$.
Substituting $v_i \rightarrow v_{\rm ss} \sim \sigma$
and $N/V = \rho_d \sim \sigma^2$ gives
$\dot{\epsilon}^{\rm pl} \sim \sigma^3$
or the natural creep exponent $m=3$. 
Larger stress exponents $m \simeq 4-6$ are often observed even in
pure metals, and are generally believed to be associated with
significantly more complex processes than those underlying natural creep.

Atomistically-informed kinetic Monte Carlo (kMC) simulations
have been used to study dislocation climb and power law creep 
in highly deformed bcc Fe \cite{kabirclimbKMC10}.
Steady state climb velocities computed from such simulations
were found to be highly nonlinear in stress,
increasing as $v_{ss} \sim \sigma^q$ with $q \simeq 3-3.5$
for large dislocation densities, $\rho_d \sim 10^{15}/m^2$-$10^{17}/m^2$.
This strong nonlinearity 
was found to arise from enhanced vacancy
diffusion rates at large $\rho_d$ as well as from the assumed vacancy 
supersaturation, applicable to highly deformed metals.
By then applying Orowan's equation,
$\dot{\epsilon}^{\rm pl} = \rho_d b v_{ss}$,
and assuming that Taylor's relation holds
in such systems, $\rho_d \sim \sigma^2$,
the authors estimated power law creep exponents directly from the measured
climb velocities as $m \simeq 5-5.5$.
This result essentially describes a modified form of natural creep,
wherein large $\rho_d$ and a steady supersaturation of vacancies
increase the `inherent' dislocation climb stress exponent from $q=1$ to 
$q \gtrsim 3$.

Clouet \cite{clouetclimbPRB11} has connected the atomistic modeling of 
Ref. [\onlinecite{kabirclimbKMC10}]
to classical mesoscale descriptions of climb, and found that extrapolation
of their results to more physically relevant dislocation densities
leads to stress exponents $m \simeq 3$, as generally expected for 
climb-assisted natural creep in metals.
The additional effects of glide, at minimum, must also apparently be 
considered if experimental stress exponents 
$m \gtrsim 4.5$ are to be reproduced.

Other mesoscale descriptions have been employed to study creep
deformation. 
Recent discrete dislocation dynamics (DDD) simulations
that incorporate both glide and climb have reproduced 
responses consistent with both diffusional and power law creep
in fcc metals \cite{dddcreep12}.
It therefore appears that the length scales,
dislocation densities, and underlying mechanisms relevant to 
some forms of power law creep may be
accessible with this type of coarse grained description.
A difficulty lies in the absence of atomistic detail,
which necessitates the use of numerous explicit rules for dislocation
interaction scenarios. 
Guidance for these rules often comes from atomistic simulations,
but these are not always universally applicable or even well-defined, and 
only a subset of all possible interactions can realistically be 
considered.
Thus, atomistic simulations that permit basic characterization
of diffusion-mediated defect evolution processes
should contribute to better informed mesoscale simulations.
Long time scale atomistic methodologies
that can access mesoscopic length scales through large scale
simulations or systematic coarse graining procedures
may also eventually provide fully self-contained 
multi-scale 
descriptions
of creep and plasticity phenomena.

\section{PFC Approach}
\label{sec:pfc}

PFC models contain atomistic detail but describe time scales in crystals
near and beyond that of the characteristic vacancy diffusion time
\cite{pfc02,pfc04,pfcdft07,pfcbook2011,pfcreview12}.
Both conservative and nonconservative 
dislocation evolution processes are captured for arbitrary crystal structure,
orientation, morphology, and applied stress
\cite{pfcdisloc06,mpfc,mpfc09,pfcavalanches10,pfcdefects12,
pfcunifieddefects2014,pfchexdef10}.
PFC simulations of diffusion-accommodated plastic flow under creep conditions
may therefore reveal previously inaccessible information about 
the atomistic mechanisms that control diffusional and 
power law creep.

A general PFC free energy functional can be written
\begin{eqnarray}
\tilde{F}
&=&\int d\vec{r}~\left[
\frac{1}{2}n^2(\vec{r})
-\frac{w}{6}
n^3(\vec{r}) + 
\frac{u}{12}
n^4(\vec{r}) \right]-\nonumber \\
& &\frac{1}{2} \int\int d\vec{r}~d\vec{r}_2~n(\vec{r})
C_2(|\vec{r}-\vec{r}_2|) n(\vec{r}_2).
\label{pfcfree}
\end{eqnarray}
where $\tilde{F}=F/(k_B T \liqr)$, 
$k_B$ is Boltzmann's constant, 
$\liqr$ is a constant reference density,
$n(\vec{r})=\rho(\vec{r})/\liqr-1$ is the scaled time-averaged
atomic density field,
$\rho(\vec{r})$ is the unscaled time-averaged atomic number density field,
$w$ and $u$ are free coefficients,
and $C_2(|\vec{r}-\vec{r}_2|)$
is the two-point direct correlation function of the fluid, assumed isotropic.
$n(\vec{r})$ may assume nonzero average values $n_0$.
We utilize the structural or XPFC class of functionals
\cite{nikpfcstruct10,nikpfcstruct11}, 
for which the Fourier transformed correlation function can be written
\be
\hat{C}_2(k)_i = 
e^{-(k-k_i)^2/(2\alpha_i^2)}
e^{-\sigma_T^2 k_i^2/(2\rho_i \beta_i)},
\label{xpfckernel}
\ee
where $i$ denotes a family of lattice planes at wavenumber $k_i$.
The constants $\alpha_i$, $\rho_i$, and $\beta_i$ are the
Gaussian width (which sets the elastic constants), planar atomic density,
and number of planes, respectively,  
associated with the $i$th family of lattice planes.
$\sigma_T$ is a Debye-Waller-type temperature parameter
that modulates the scattering intensity 
($S(k)=(1-\hat{C}_2(k))^{-1}$ where $S(k)$ is the structure factor)
due to the effect of atomic thermal vibrations.
The envelope of all selected Gaussians $i$ 
composes the final $\hat{C}_2(k)$.
A single reflection 
\cite{footnote1}
at $k_1=2\sqrt{3}\pi$ is used here
to produce equilibrium bcc structures with lattice constant 
$a \simeq \sqrt{2/3}$.

The standard stochastic nonlinear diffusion equation is employed
for $n(\vec{r})$ dynamics,
\be
\frac{\partial n(\vec{r})}{\partial \bar{t}} = \nabla^2\frac{\delta 
\tilde{F}}
{\delta n(\vec{r})}
+\eta(\vec{r},\bar{t}),
\label{eq:pfcdyn1}
\ee
where $\bar{t}=\Gamma t/\liqr^2$ 
is rescaled time (denoted as $t$ in subsequent sections),
$\Gamma$ is a mobility constant, 
and $\eta(\vec{r},\bar{t})$ is a Gaussian stochastic noise variable with 
$\langle\eta(\vec{r},\bar{t})\rangle=0$ and
$\langle\eta(\vec{r}_1,\bar{t}_1)\eta(\vec{r}_2,\bar{t}_2)\rangle
=-2 k_B T \nabla^2\delta(\vec{r}_1-\vec{r}_2)\delta(\bar{t}_1-\bar{t}_2)$.
PFC studies of mechanical properties often employ an equation of motion
with an additional inertial or wave-like term,
$\partial^2 n / \partial \bar{t}^2$, to facilitate rapid elastic
relaxations \cite{mpfc}.
Since the diffusive component of the mechanical response is our primary 
interest in this work, and since we focus on the low stress and low
steady-state strain rate regime in which inertial effects should be secondary,
we employ the simpler diffusive form of Eq.\ (\ref{eq:pfcdyn1}). 
No qualitative differences in terms of the creep-type response outlined in
the following sections were observed in comparative simulations 
employing inertial dynamics.

We further define $M=2 k_B T$ and use this variable 
in place of $k_B T$ to specify a 
scaled temperature. 
This description thus contains two explicit temperature parameters,
$\sigma_T$ and $M$, as well as any implicit temperature dependences
of the other parameters in Eqs.\ (\ref{pfcfree}) and (\ref{xpfckernel}).
For simplicity and following convention, it is assumed that
any implicit $T$ dependences in these latter scaled parameters can be 
neglected to lowest order.
$M$ and $\sigma_T$ can thus be varied simultaneously to control
the physical temperature $T$
in the vicinity of the melting temperature ($T_m$),
though either parameter has the same basic effect independently for our 
purposes. For simplicity, we therefore control $T$ by varying only $M$ 
while holding $\sigma_T$ fixed, unless noted otherwise.
All simulations in this study were performed in 3D using a
pseudo-spectral algorithm
with semi-implicit time stepping
and periodic boundary conditions.

Within standard PFC descriptions, the explicit evolution of vacancies is
interpreted as being coarse grained into the structure of $n(\vec{r})$ and
its diffusive evolution.
One question that we wish to address is whether this coarse grained description
of vacancy diffusion can reproduce Nabarro-Herring and Coble creep 
mechanisms, which involve only the strain-induced flow of vacancies.

It is also currently unclear, in terms of power law creep, whether
the length scales, dislocation densities, and collective defect kinetics
relevant to $m \gtrsim 3$ kinetics can be reached by PFC-type models.
The case of natural creep, as outlined in Section \ref{sec:creep},
provides an illustration of this issue.
Since the kinetics of climb in PFC models have been shown to follow
$v_{\rm ss} \sim \sigma$ for small
dislocation density \cite{pfcdisloc06},
it stands 
to reason
that a realization of Taylor's relation would lead directly to $m=3$
natural creep, 
or potentially $m=4.5$ creep when glide is also considered \cite{weertman75}.
It has not yet been demonstrated that
the system sizes and dislocation creation-annihilation kinetics
needed to self-consistently reproduce 
phenomena such as 
Taylor's relation can be simulated
by PFC models.
Though we do not attempt to simulate large 3D systems
with multiple slip systems in which Taylor's relation generally
emerges, 
analagous behaviors in quasi-2D systems are examined further in the following.

We numerically investigate the response of bcc crystals and polycrystals
to creep-type deformation as follows.
As in typical creep experiments,
deformation is applied under constant stress conditions, using a
newly developed stress-controlled PFC simulation method.
At each evolution iteration $\Delta t$, the grid spacing $\Delta_i$
is homogeneously varied in one or more directions $i$,
depending on the particular choice of boundary constraints,
until a specified stress $\sigma_{A}$ is achieved.
We consider quasi-2D system geometries 
with columnar grains aligned along the $x$ axis and strain $\epsilon$ applied 
along the $y$ and $z$ axes (see Fig.\ \ref{hexpics1}),
and further employ a
fixed $y$-$z$ area constraint 
\cite{footnote2}
with $\sigma_{xx}=0$.
$\Delta_z$ and $\Delta_y$ are simultaneously varied as
$\Delta_z=\Delta_z^0(1+\epsilon)$ and
$\Delta_y=\Delta_y^0/(1+\epsilon)$,
where $\Delta_i^0$ is the initial grid spacing along $i$,
until the condition 
$\partial\tilde{f}/\partial\epsilon = \sigma_{A}$
is achieved, where $\tilde{f}=\tilde{F}/V$ is the mean total free energy.
$\Delta_x$ is then similarly varied until
$\partial\tilde{f}/\partial\Delta_x = 0$,
though generally $\Delta_x \simeq \Delta_x^0$ at all times
for the columnar geometry employed. 
This method is therefore effectively equivalent to a constant volume 
pure shear deformation
with fixed $\sigma_{A}=\sigma_{yy}+\sigma_{zz}$.

We have confirmed that the total strain generated 
versus $\sigma_{zz}$
for the case $\sigma_{xx}=\sigma_{yy}=0$
reproduces the correct linear stress-strain relation of the perfect
one-mode bcc crystal for $\epsilon \lesssim 0.02$.
As expected,
modest deviations from linearity are observed for $\epsilon \gtrsim 0.02$
when nonlinear elastic effects become appreciable.
We have also verified that the one-mode bcc Poisson ratio $\nu=1/3$ is
realized at low $\sigma_{zz}$ in these simulations,
again with nonlinear effects causing a modest gradual decrease to 
$\nu \simeq 0.31$ by $\epsilon = 0.08$.

\section{Diffusional Creep and
Stress-Assisted Grain Boundary Migration in Nanopolycrystals}
\label{sec:hexcreep}

\begin{figure}[btp]
 \centering{
 \includegraphics*[width=0.48\textwidth,trim=0 0 0 0]{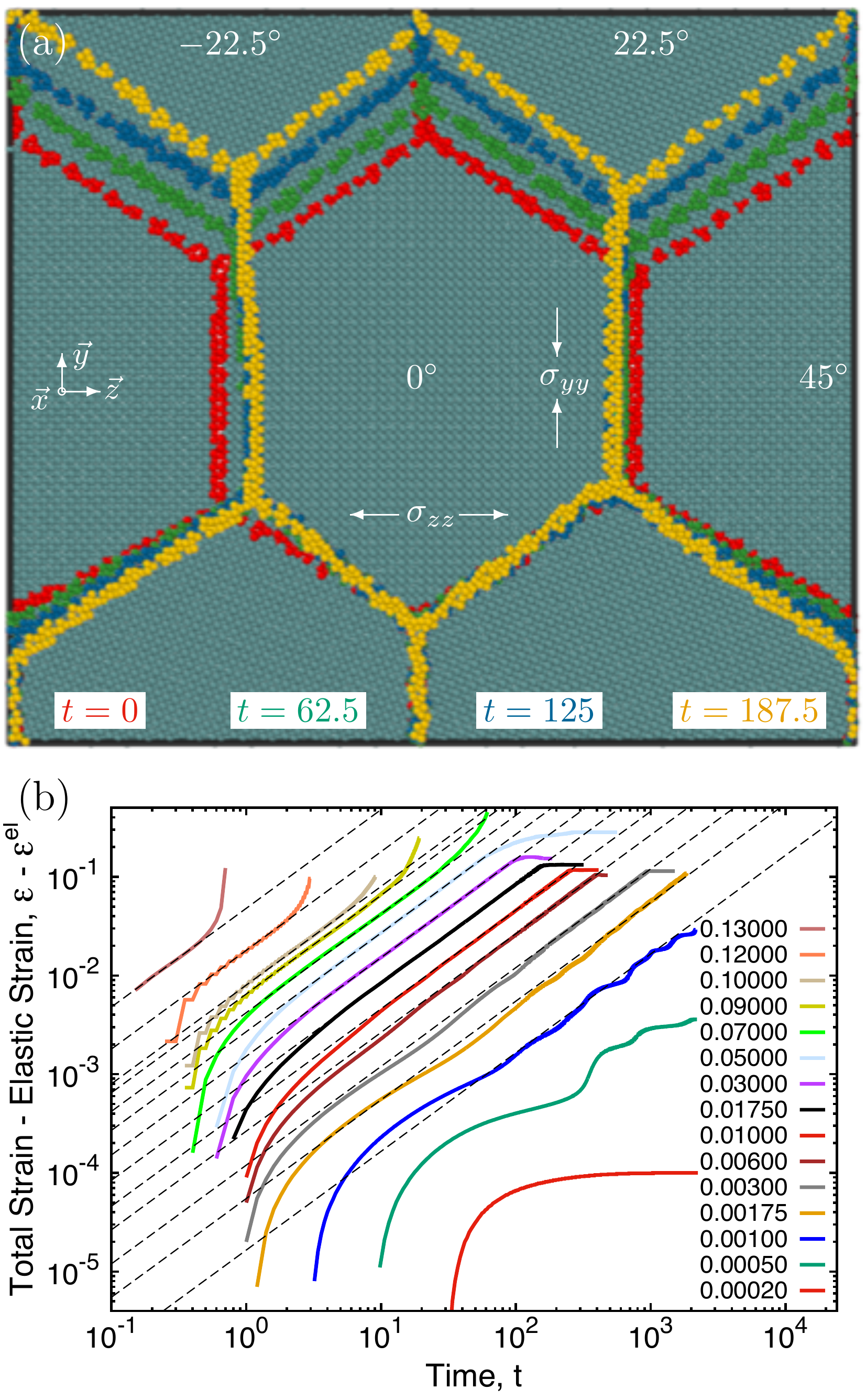}
 }
 \vspace{-.5cm}
\caption[]
{\label{hexpics1}
(Color online)
Diffusional creep in bcc nanopolycrystals with high angle grain boundaries.
(a) Strain normalized system configurations for $\sigma_A=0.01$ and $d=50.8a$
are shown at
($0t$, $\epsilon=0.0$), 
($62.5t$, $\epsilon=0.031$), 
($125t$, $\epsilon=0.060$), and
($187.5t$, $\epsilon=0.090$).
The system thickness is $L_x=1a$.
Strain normalization is given by 
$y \rightarrow y(1+\epsilon)$ and
$z \rightarrow z/(1+\epsilon)$.
For analysis and visualization purposes, local peaks in 
$n(\vec{r})$, which represent the most probable atomic positions, 
are taken to correspond to atomic sites.
Sites with bcc coordination (at $t=0$) are shown in pale green, 
those with irregular
coordination (grain boundary atoms) are shown in 
red ($t=0$), green ($t=62.5$), pale blue ($t=125$), and gold ($t=187.5$).
(b) A representative set of creep curves for the configuration of 
(a).
$\epsilon^{\rm pl}$ vs.\ time is shown at various $\sigma_A$
(color legend).
The dashed black lines are linear fits.
}
\end{figure}

\begin{figure}[btp]
 \centering{
 \includegraphics*[width=0.48\textwidth,trim=0 0 0 0]{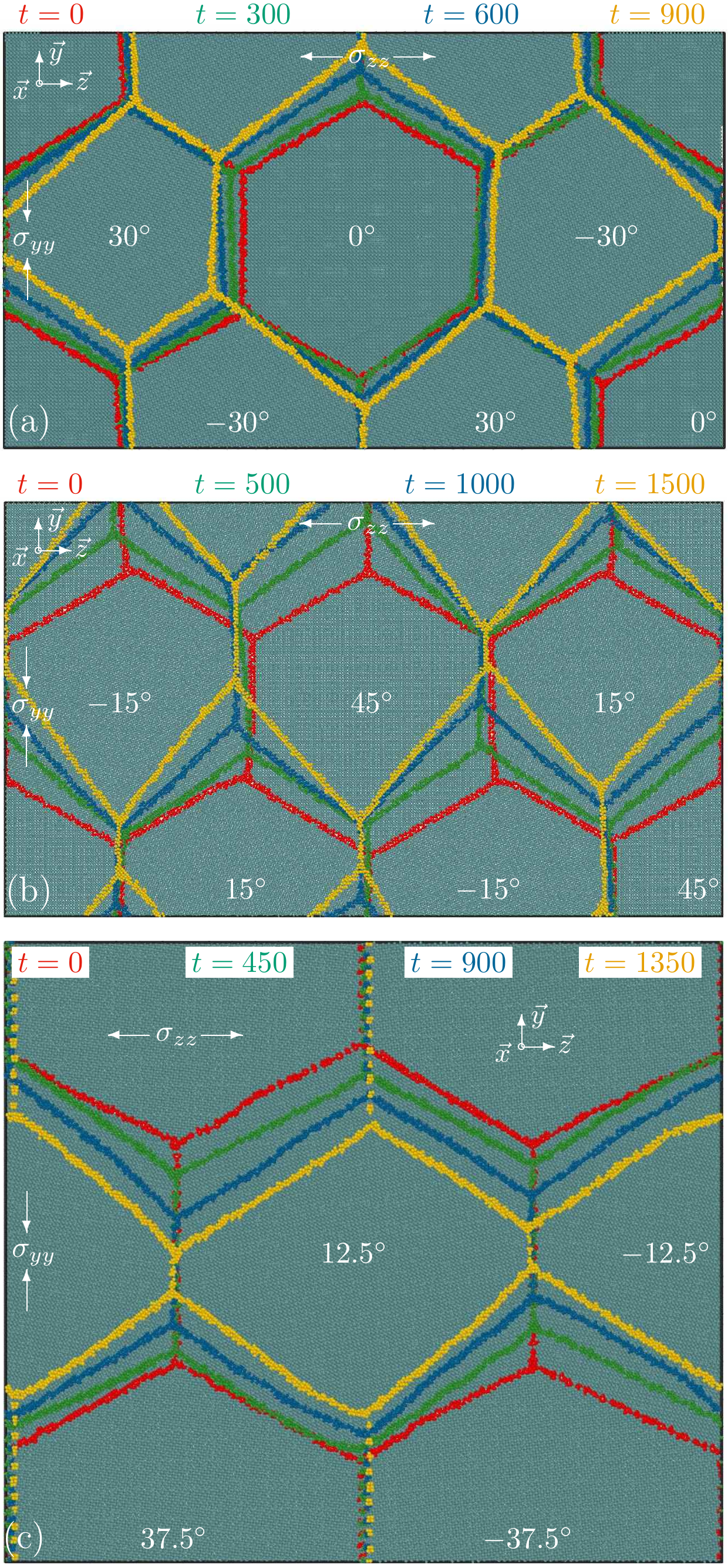}
 }
 \vspace{-.5cm}
\caption[]
{\label{hexpics2}
(Color online)
Morphology of diffusional creep in other bcc nanopolycrystal configurations
with high angle grain boundaries.
Strain normalized system configurations are shown for:
(a) The symmetric 6 grain cell with rotations of ($0^{\circ}$, $\pm 30^{\circ}$)
and $\sigma_A=0.01$, $d=76.3a$. 
Red: ($0t$, $\epsilon=0.0$), 
green: ($300t$, $\epsilon=0.044$), 
pale blue: ($600t$, $\epsilon=0.090$),
gold: ($900t$, $\epsilon=0.135$).
(b) The symmetric 6 grain cell with rotations of ($45^{\circ}$, $\pm 15^{\circ}$)
and $\sigma_A=0.01$, $d=76.3a$. 
Red: ($0t$, $\epsilon=0.0$), 
green: ($500t$, $\epsilon=0.108$), 
pale blue: ($1000t$, $\epsilon=0.234$),
gold: ($1500t$, $\epsilon=0.384$).
(c) The asymmetric 4 grain cell with rotations of 
($\pm 12.5^{\circ}$, $\pm 37.5^{\circ}$)
and $\sigma_A=0.01$, $d=101.7a$.
Red: ($0t$, $\epsilon=0.0$), 
green: ($450t$, $\epsilon=0.055$), 
pale blue: ($900t$, $\epsilon=0.111$),
gold: ($1350t$, $\epsilon=0.166$).
}
\end{figure}

An idealized columnar grain structure with initially uniform grain shapes
and sizes was used to examine diffusional creep in bcc nanopolycrystals.
Periodic systems with either 4 or 6 hexagonal grains per repeat unit
were chosen, with various symmetric and asymmetric 
combinations of grain rotations and grain boundary tilt angles.
High angle boundaries were examined in
asymmetric 4 grain systems
with grain rotations of ($0^{\circ}$, $45^{\circ}$, $\pm 22.5^{\circ}$) or
($\pm 12.5^{\circ}$, $\pm 37.5^{\circ}$). 
The former configuration contains
four $45^{\circ}$ and eight $22.5^{\circ}$ boundaries, while the latter
contains two $15^{\circ}$, two $25^{\circ}$, and eight $40^{\circ}$ boundaries, as shown in Figs.\ \ref{hexpics1}(a) and \ref{hexpics2}(c), respectively.
Symmetric 6 grain systems
with grain rotations of ($0^{\circ}$, $\pm 30^{\circ}$) or
($45^{\circ}$, $\pm 15^{\circ}$) were also examined.
These configurations contain eighteen $30^{\circ}$ boundaries, 
as shown in Figs.\ \ref{hexpics2}(a) and \ref{hexpics2}(b).
Low angle grain boundaries were examined using a similar asymmetric 6 grain
configuration with grain rotations of ($0^{\circ}$, $\pm 5^{\circ}$).
This system contains
twelve $5^{\circ}$ boundaries and six $10^{\circ}$ boundaries
(see Fig.\ \ref{hexlagb}(a) ahead). 
The columnar axis is along $\vec{x}=[100]$ in all cases,
and a system thickness $L_x=1a$ was used in this direction
(no changes were observed for $L_x=10a$).
Grain sizes $d=25.4a$, $50.8a$, $76.3a$, $89.0a$, $101.7a$, $114.4a$, $127.1a$, 
and $152.5a$ were studied,
where $d$ is the diameter of the smallest circle that encloses the hexagon.
Unless specified otherwise, model parameters
$w=1.4$, $u=1$, $n_0=0$, $\alpha_1=1$,
$\sigma_T=0.1$, $\rho_1=1$, $\beta_1=8$, and $M=0$
are used with 
$\Delta_i^0 \simeq a/12$
and $\Delta t=0.01$,
where $a=0.81675$.

After equilibration of each system at 
$\sigma_{xx}=\sigma_{yy}=\sigma_{zz}=0$,
simulations were
conducted at a range of $\sigma_A$ values,
and the subsequent deformation $\epsilon(t)=[L_z(t)-L_z(0)]/L_z(0)$ 
was monitored as a function of time.
Analysis of creep dynamics is restricted to stresses above 
$\sigma_{-}$,
the threshold for observable plasticity 
($\sigma_{-} \simeq 0.0006$) 
and below 
$\sigma_{+}$,
the threshold
for dislocation nucleation from grain boundaries
($\sigma_{+} \simeq 0.06$). 
The absence of lattice dislocations or other defects in the grain interiors
allows us to isolate and characterize grain boundary-mediated
plasticity mechanisms 
without complications from collective dislocation processes, for example.
Plastic flow in these grain geometries generally requires a significant
amount of nonconservative 
grain boundary motion, i.e., vacancy-mediated migration and climb.

Even though all grains are initially identically hexagonal, their
different orientations and boundary structures create small asymmetries 
upon equilibration, which grow when stress is applied.
The result is an initial delta function
grain size distribution that broadens with time 
due to stress-assisted grain boundary migration or grain growth.
The symmetric 6 grain geometry described above has been used in MD to 
avoid precisely this effect \cite{creepMillett2008}, but 
it will be shown in the following that 
the mobility of grain boundaries in our PFC simulations is large
enough to significantly impact the measured steady-state creep rates.
In discussing evolution of the grain size distribution, 
we distinguish between net grain growth, an increase
in the average grain size, and differential growth, a broadening of the
grain size distribution without any increase in average size.
Net growth is observed only at very late times when entire grains are
eliminated, after the steady-state creep regime has ended,
and therefore does not influence our results. 
Differential growth does occur during the steady-state creep regime.

\subsection{High angle grain boundaries}
\label{subsec:hagb}
Representative creep curves obtained for 
the asymmetric 4 grain system
with ($0^{\circ}$, $45^{\circ}$, $\pm 22.5^{\circ}$) rotations
are shown in 
Fig.\ \ref{hexpics1}(b)
for $d=50.8a$.
After an initial elastic strain $\epsilon^{\rm el}$ 
and a brief regime of plastic flow onset
(primary creep), an extended regime of linear steady-state creep
(secondary creep) is observed for all $\sigma_- < \sigma_A < \sigma_+$.
The slope of the best linear fit to this regime gives the steady-state
creep rate $\dot{\epsilon}^{\rm ss}$ for a given $\sigma_A$.
The steady-state creep regime terminates in one of two ways.
For $\sigma_A \lesssim 0.06$, 
differential grain growth proceeds until
two grains eventually consume the others at late times (net grain growth),
leaving two $45^{\circ}$ boundaries along $y$
that no longer migrate in response to $\sigma_A$.
Creep therefore ceases and all strain energy is absorbed elastically.
For $\sigma_A \gtrsim 0.06$, dislocations eventually nucleate from the grain
boundaries or homogeneously from the grain interiors at late times.
This leads
to unbounded plastic flow, a divergence in the strain rate, and loss of
mechanical integrity.

\begin{figure}[btp]
 \centering{
 \includegraphics*[width=0.48\textwidth,trim=0 0 0 0]{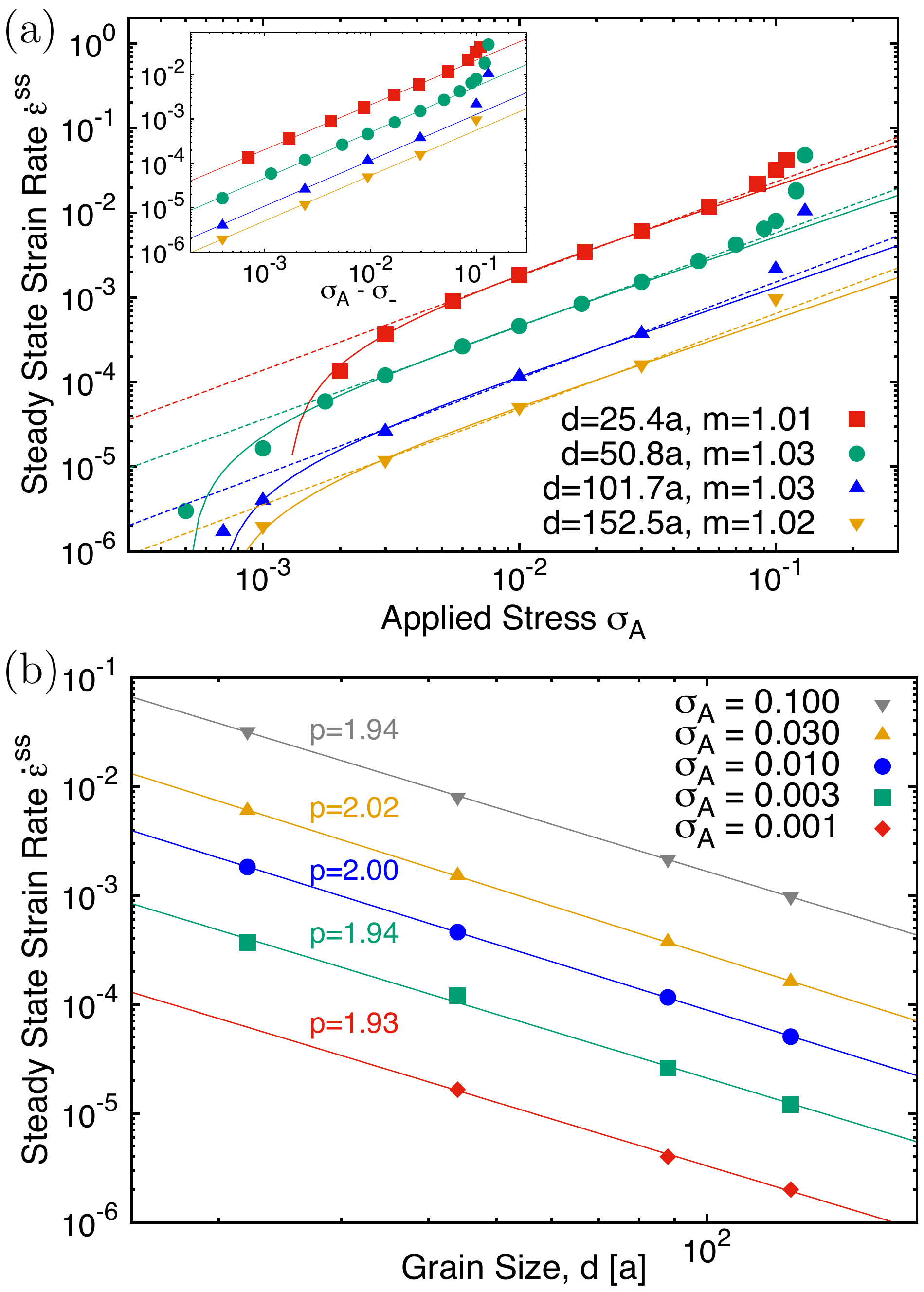}
 }
 \vspace{-.25cm}
\caption[]
{\label{hexdata2}
(Color online)
Creep exponents of the configuration shown in Fig.\ \ref{hexpics1}.
(a) Compilation of steady-state creep rates vs.\ $\sigma_A$ 
at various grain sizes $d$.
The dashed lines are fits to Eq.\ (\ref{creepeq}),
the solid lines are fits to Eq.\ (\ref{creepeq}) with
$\sigma_A$ replaced by $\sigma_A - \sigma_-$.
The resulting creep rate
stress exponents $m$ indicated in the figure
are those of the solid lines, while
the corresponding values for the dashed lines are
$m=1.11$, $1.10$, $1.14$, and $1.13$, respectively.
The inset shows the same data plotted vs.\
$\sigma_A - \sigma_-$ with the same solid line fits.
(b) Dependence of $\dot{\epsilon}^{\rm ss}$ on grain size $d$ at various
$\sigma_A$.
Solid lines are fits to Eq.\ (\ref{creepeq}) with variable grain size
exponent $p$. 
See Supplemental Material 
\cite{footnote3}
for animations of the 4 simulations at $\sigma_A=0.01$.
}
\end{figure}

The extracted steady-state creep rates $\dot{\epsilon}^{\rm ss}$ for four
values of $d$ are plotted as a function of $\sigma_A$ in 
Fig.\ \ref{hexdata2}(a).
The best fit slopes of these data
for $\sigma_- < \sigma_A < \sigma_+$ (dashed lines)
indicate that the bare stress exponents are 
$m = 1.12 \pm 0.02$. 
Taking the effect of $\sigma_-$ into account by substituting
$\sigma_A-\sigma_-$ for $\sigma_A$ produces
$m = 1.02 \pm 0.02$ (solid lines and inset).
The dependence of
$\dot{\epsilon}^{\rm ss}$ on $d$ 
for 5 representative values of $\sigma_A$ 
is plotted in Fig.\ \ref{hexdata2}(b).
All data sets are well fit by a grain size exponent 
$p = 1.98 \pm 0.05$.

Very similar values for the exponents $m$ and $p$ are obtained
for the other high angle grain boundary configurations shown in
Fig.\ \ref{hexpics2}.
The morphology of the grain structures, on the other hand, 
does vary with the details of the initial configuration.
For example, the initial systems shown in
Figs.\ \ref{hexpics2}(a) and \ref{hexpics2}(b) are identical
except for a rotation of all grains by $45^{\circ}$ about the
$\vec{x}$ axis, but their structures evolve quite differently. 
Whereas the two $0^{\circ}$ grains in Fig.\ \ref{hexpics2}(a)
grow nearly radially while the $\pm 30^{\circ}$ grains shrink, 
all six grains in Fig.\ \ref{hexpics2}(b)
elongate and coherently translate nearly uniformly along $\vec{y}$.
In the asymmetric 4 grain systems shown in Figs.\ \ref{hexpics1}(a) and 
\ref{hexpics2}(c), a relatively fixed $30^{\circ}$ orientation is
maintained along the boundaries that do not lie parallel to $\vec{y}$.
These boundaries translate nearly uniformly along $\vec{y}$
until two of the four grains are eliminated.
The causes of these differences are examined further at the end of this
section.
We also note that grain rotations are observed in coordination with
differential grain growth, and are a consequence of nonconservative
boundary migration.

\begin{figure}[btp]
 \centering{
 \includegraphics*[width=0.48\textwidth,trim=0 0 0 0]{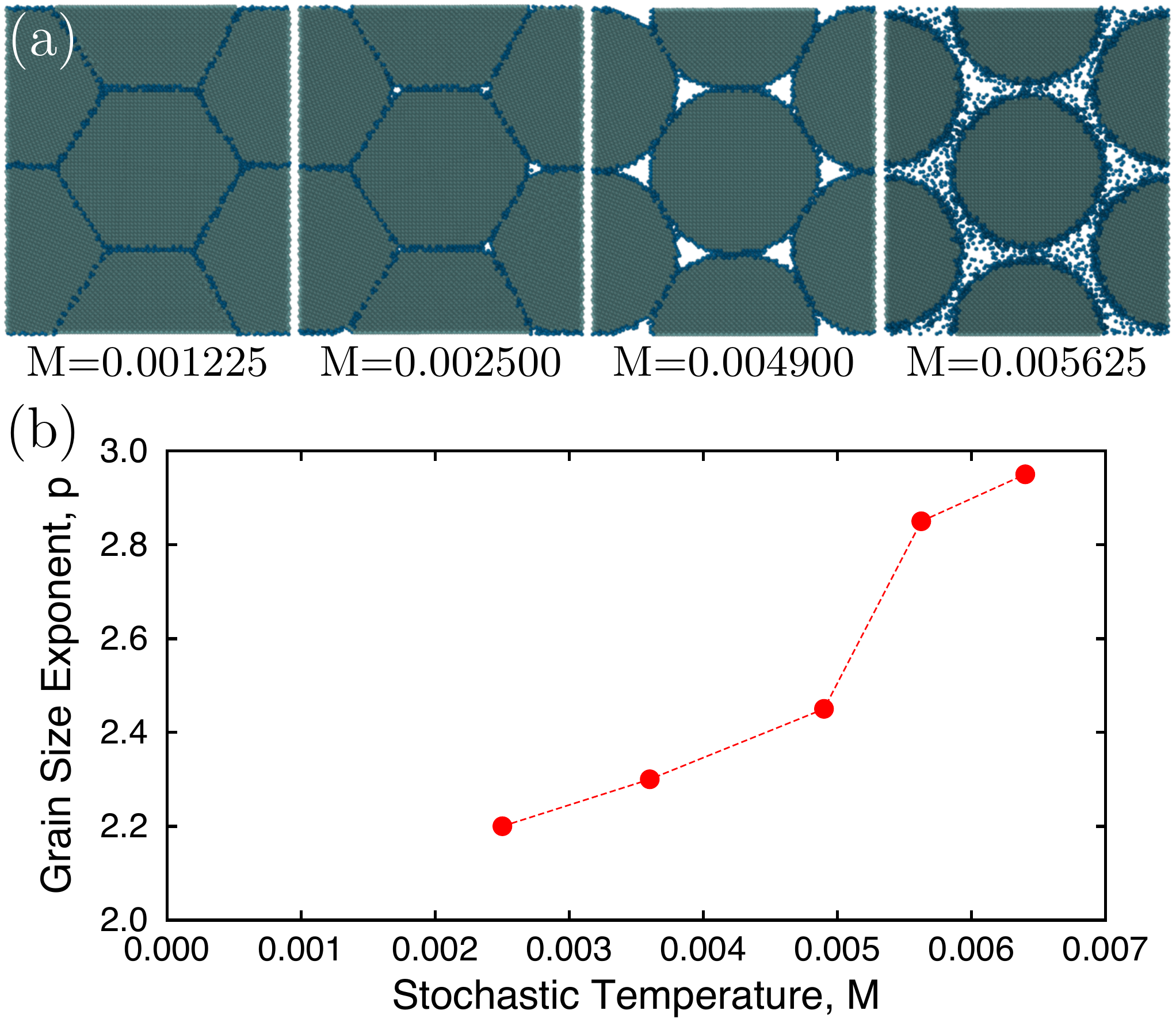}
 }
 \vspace{-.5cm}
\caption[]
{\label{premelt}
(Color online)
(a) Effect of stochastic temperature $M \sim k_B T$ on 
grain boundary structure.
Colors are as in Fig.\ \ref{hexpics1}(a).
(b) Variation of the grain size exponent $p$ with $M$.
}
\end{figure}

Interestingly, we observe a transition from
$m \simeq 1$ and $p \simeq 2$ behavior to 
$m \simeq 1$ and $p \simeq 3$
as the stochastic temperature $M$
is raised to very near melting.
The effects of $M$ on grain boundary structure and the corresponding
grain size exponents $p$ are shown in Fig.\ \ref{premelt}.
There is a clear correlation between the increase in $p$ from $\sim$2
to $\sim$3 and the effective liquification of the grain boundaries.
The observed exponents are consistent with those of Coble creep, which
is typically dominant only at relatively low $T$.
Our findings therefore indicate that
grain boundary premelting in the presence of low applied stresses
can result in a re-entrant Coble creep regime at $T$ sufficiently
near the equilibrium melting point.

\subsection{Low angle grain boundaries}
\label{subsec:lagb}

\begin{figure}[btp]
 \centering{
 \includegraphics*[width=0.48\textwidth,trim=0 0 0 0]{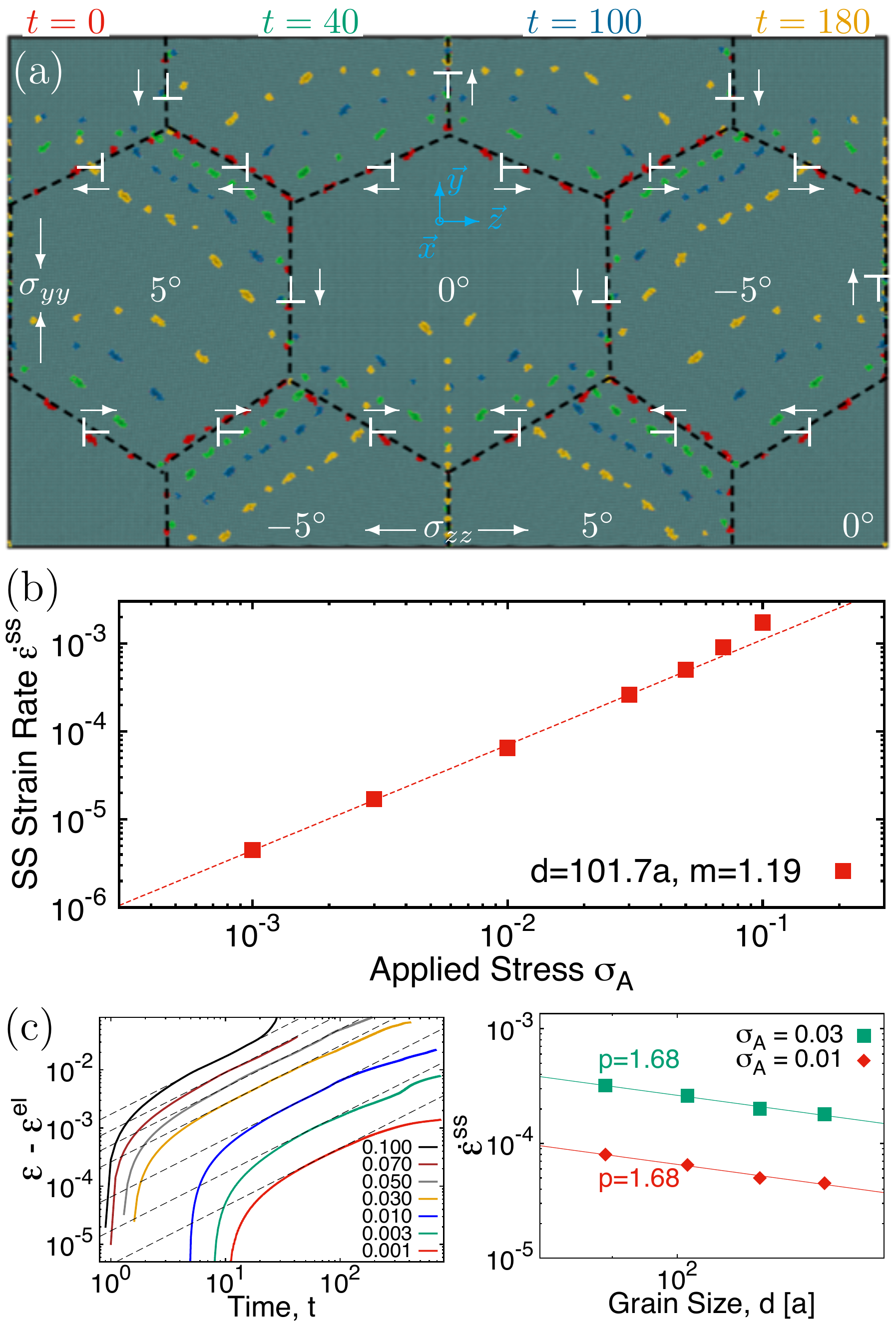}
 }
 \vspace{-.5cm}
\caption[]
{\label{hexlagb}
(Color online)
Diffusional creep in bcc nanopolycrystals with only
low angle grain boundaries.
(a) Strain normalized
system configurations for $\sigma_A=0.03$ and $d=101.7a$ are shown at
($0t$, $\epsilon=0.0$), 
($40t$, $\epsilon=0.0211$), 
($100t$, $\epsilon=0.0342$), and
($180t$, $\epsilon=0.0499$).
The system thickness is $L_x=1a$.
Sites with bcc coordination (at $t=0$) are shown in pale green, 
those with irregular coordination (dislocation core atoms) are shown in 
red ($t=0$), green ($t=40$), pale blue ($t=100$), and gold ($t=180$).
Burgers vectors of dislocations along each boundary are indicated with
white $\perp$ symbols, along with the subsequent climb directions
expected upon deformation.
(b) Steady state creep rates vs.\ $\sigma_A$ at $d=101.7a$.
The dashed line is a fit to Eq.\ (\ref{creepeq}) with $m=1.19$.
(c) Left: Creep curves for the data shown in (b).
$\epsilon^{\rm pl}$ vs.\ time is shown at various $\sigma_A$
(color legend).
The dashed black lines are linear fits.
Right: 
Dependence of $\dot{\epsilon}^{\rm ss}$ on grain size $d$ at
$\sigma_A=0.01$ and $0.03$.
Solid lines are fits to Eq.\ (\ref{creepeq}) with grain size
exponent $p=1.68$. 
}
\end{figure}

The 6 grain low angle grain boundary configuration evolves as
shown in Fig.\ \ref{hexlagb}(a).
Since the Burgers vectors of the grain boundary dislocations can
be clearly identified, the direction of the driving force for each boundary
under $\sigma_{zz}$ and $\sigma_{yy}$ is readily known
(white arrows in Fig.\ \ref{hexlagb}(a)).
Indeed, the motion of each boundary is consistent with that expected
from its Burgers vector.
Deviations from the exact direction expected are 
due to forces exerted by neighboring boundaries,
as discussed further in the following subsection.
Motion of the dislocations within boundaries aligned with 
$\vec{y}$ occurs purely by climb,
while motion of all other boundary dislocations occurs via a mixture
of climb and glide.
The stress exponent $m$ obtained during the regime of morphologies
roughly corresponding to those shown in Fig.\ \ref{hexlagb}(a) 
is $m \simeq 1.19$
(Fig.\ \ref{hexlagb}(b)).
Eventually, boundaries begin to meet and both the dislocation density
and strain rate decrease at later times as dislocations annihilate.
Figure \ref{hexlagb}(c) shows the grain size dependence of
$\dot{\epsilon}^{\rm ss}$, from which an exponent $p \simeq 1.68$
is obtained.

\subsection{Discussion}
\label{subsec:hexdiscussion}

The creep exponents $m \simeq 1$ and $p \simeq 2$ obtained
in the preceeding simulations are consistent with
those of Nabarro-Herring creep, 
grain redistribution mediated by bulk vacancy diffusion.
As discussed in Section \ref{sec:creep},
idealized diffusional creep would be expected to generate
a nearly affine deformation morphology,
such that no significant structural changes would be observed upon normalization
of the system configuration by strain
($y \rightarrow y(1+\epsilon)$, $z \rightarrow z/(1+\epsilon)$),
no matter the initial grain orientations and patterns.
This is clearly not the case in any of the present simulations.
The grain deformation seen in Fig.\ \ref{hexpics2}(b) is the closest to
affine, but some elongation along $\vec{y}$ is observed as well a
significant amount of collective translation in this direction.
During the primary and steady-state deformation regimes, 
the summed grain boundary length and the number of grains 
within each system remain fixed,
but the grain size distributions broaden.
Differential grain growth or stress-assisted grain boundary migration
therefore clearly influences the measured creep rates.
In this subsection, we attempt to characterize the relative degrees to which
Nabarro-Herring creep and stress-assisted grain boundary migration
mediate plastic flow in these simulations, and to explore connections,
similarities, and differences between these two related mechanisms.

The low angle grain boundary simulations are perhaps most easily understood
from the viewpoint of stress-assisted migration.
The expected climb direction of each dislocation/boundary,
based on its Burgers vector and the direction of applied stress,
is indicated by a white arrow in Fig.\ \ref{hexlagb}(a).
The general agreement with the observed migration directions
indicates that the applied stress directly induces dislocation climb
and therefore boundary migration.
The entire structure then evolves collectively in response to these
climb forces, while
the topological constraints of the boundary network transmit
additional network forces to each boundary.
For example, the dislocations within the upper left boundary of the
central $0^{\circ}$ grain migrate in the -$\vec{z}$ direction
due to the climb force generated by $\sigma_{yy}$,
but they also migrate in the +$\vec{y}$ direction due to the motion
of the neighboring vertical boundary in this direction.
The net result is a mixed climb-glide migration nearly along
the vector $\hat{y}-\hat{z}$.
The two $10^{\circ}$ vertical boundaries translate only along
$\vec{y}$ by pure climb, as the forces due to neighboring boundaries
balance along $\vec{z}$.
The four $5^{\circ}$ vertical boundaries climb similarly along $\vec{y}$,
but also experience a net torque about $\vec{x}$ due to the opposing
migration directions of their neighbors along $\vec{y}$ and -$\vec{y}$,
respectively. 

By tracking the evolution of individual dislocations, we have quantified the
amount of plastic strain relief due to dislocation motion.
We find that this mode of strain relief accounts for nearly all
of the applied system-wide strain, confirming that dislocation/boundary
migration
is the primary mechanism of plasticity in this case.

Though high angle grain boundaries cannot be analyzed in terms
of individual dislocations, the relatively
similar differential grain growth morphologies
and creep exponents observed across low and high angle systems suggest
that no qualitative change in plasticity mechanisms occurs with angle.
The probability of 
conservative grain boundary sliding should nonetheless
become more significant as the boundary angles increase.
Visual inspection of $n(\vec{r})$ during high angle boundary migration
reveals that density peaks are created 
within boundaries under tension (positive mass flux) and destroyed
within boundaries under compression (negative mass flux),
while no signs of significant boundary sliding have been observed.
This flow of mass is consistent with the basic Nabarro-Herring process.
It is therefore perhaps not surprising that 
the Nabarro-Herring exponents, $m=1$ and $p=2$, are observed
even though the grain morphology does not
remain affine during concurrent stress-assisted migration.

These findings suggest that plasticity in the present PFC simulations
is mediated by the inherently coupled processes of vacancy diffusion and 
grain boundary migration, in the following general manner.
Applied stress generically induces vacancy flow out of grain boundaries under
tension and into those under compression.
These flows should in turn be capable of driving both the grain redistribution
of diffusional creep and the nonconservative defect evolution of 
stress-assisted boundary migration.
In our simulations, diffusion apparently
occurs primarily through the grain interiors, in accord with
idealized Nabarro-Herring creep. Rather than simply depositing 
onto non-tensile boundaries such that affine grain 
redistribution and elongation occurs, 
the vacancies directly facilitate climb-mediated grain boundary migration.
This migration is possible because of the large nonconservative
mobilities of PFC dislocations and grain boundaries relative to the
inherent diffusive time scales of vacancies in this description.
The subsequent directions of grain boundary
migration are determined at the atomistic level by the 
dislocation orientations and/or grain boundary topologies.
These should generally exhibit no predisposition for affine morphologies,
as is clear for the case of Fig.\ \ref{hexlagb}(a).
The symmetry of the grain shapes is thus broken, a
distribution of sizes emerges, and differential growth
proceeds, driven by the stress-induced vacancy fluxes.
The correspondence in terms of underlying mechanism suggests that
this type of creep plasticity in the large boundary mobility regime
may reasonably be expected to produce
the same stress and grain size exponents as those of idealized
Nabarro-Herring creep, though with very different effects on grain morphology,
particularly for nanoscale grains.

The non-affine morphologies appear to emerge here largely from 
atomistic effects, many aspects of which would therefore be difficult 
to predict without atomistic insight.
First, variation of grain boundary structure with angle leads to
breaking of the perfect hexagonal symmetry under zero stress.
Second, the preferred and realized kinetics of boundary migration under stress
are determined by a number of potential effects with atomic-scale origins.
For example,
the stress dependence of the grain boundary mobilities and their preferred
migration directions are linked to grain boundary/dislocation structure.
The driving force for migration may also be influenced by grain orientation 
if elastic anisotropy is sufficiently large.
Further investigation is needed to determine which of these or other factors 
most influence the nature of this type of stress-assisted boundary migration,
particularly for high angle boundaries,
but over the range of parameters and deformation conditions investigated,
our results indicate that it
can be a significant contributor to creep flow in the regime of large
boundary mobilities.

\section{Dislocation or Power Law Creep}
\label{sec:disloccreep}

To examine the transition to dislocation-mediated plastic flow
and potentially power law creep, 
we must, at minimum,
introduce dislocation sources 
into our simulations.
Naturally emerging steady-state $\rho_d$
and $\dot{\epsilon}$ values involving primarily dislocation climb
and glide would then signal a physically meaningful power law creep regime.
In the following subsection, 
we begin by examining a simple controlled geometry in
which the number of mobile dislocations
$N$ does not depend significantly on $\sigma_A$,
and a stress exponent $m=1$ is therefore
expected (this is still technically dislocation, not diffusional, creep).
It will be shown that $m \gg 1$ can nonetheless arise due to
additional factors associated primarily with large
dislocation densities.
Other scenarios are then considered in
which $N$ does increase with $\sigma_A$ and larger stress exponents are 
thus expected, as in the case of natural creep.
Connections are drawn between the subsequent power law stress exponents
and the dislocation density effects identified in the constant $N$ case.

\subsection{Dislocation nucleation, climb, and pile-up}
\label{subsec:disloctheory}

\begin{figure*}[btp]
 \centering{
 \includegraphics*[width=0.80\textwidth,trim=0 0 0 0]{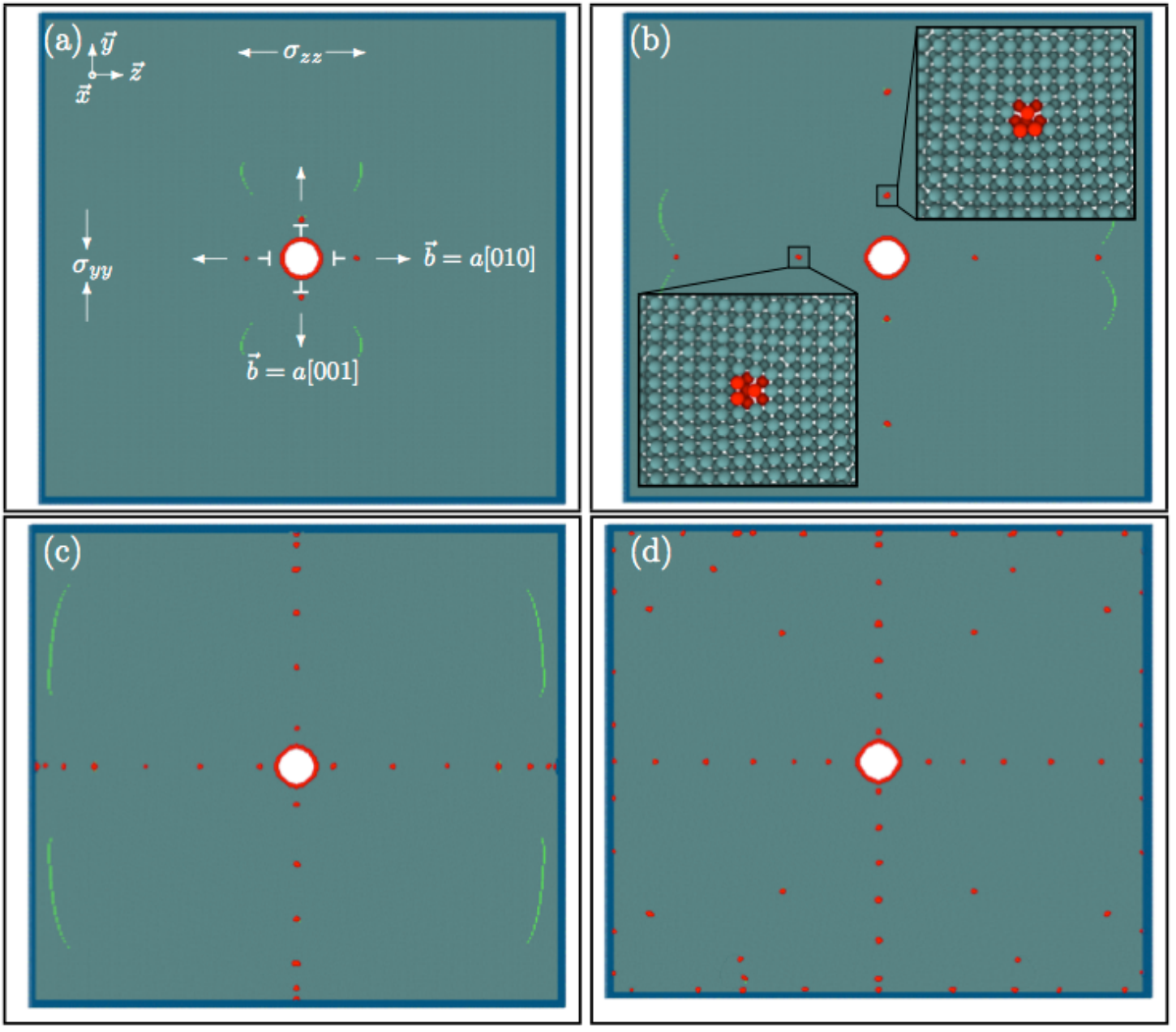}
 }
 \vspace{-.25cm}
\caption[]
{\label{wallpics}
(Color online)
Dislocation nucleation, climb, and pile-up in a simple dislocation creep
scenario at $\sigma_A=0.09$ and $L_y=L_z=264a$.
(a) $50t$, $\epsilon=0.0358$, 
(b) $130t$, $\epsilon=0.0397$, 
(c) $360t$, $\epsilon=0.0544$, 
(d) $480t$, $\epsilon=0.0751$.
Interior atoms with bcc coordination are shown in pale green, 
exterior hard wall atoms (also bcc coordination) in pale blue,
those with irregular coordination (dislocation core atoms) in red, 
and those with fcc coordination in bright green.
Red atoms are displayed with larger radii
to highlight the dislocation positions.
}
\end{figure*}

To begin examining the effects of dislocation climb on PFC creep rates, 
we first characterize a simple system geometry consisting of a perfect
bcc crystal with one central dislocation source, as shown in 
Fig. \ref{wallpics}.
The simulation cell and crystallographic directions are the same
as those of the previous section, as is the system thickness of $L_x=1a$.
The dislocation source is a cylindrical region of radius $R=10a$ in which 
$n(\vec{r})$ is slaved to a homogeneous penalty function \cite{mpfc09}.
Though the model is capable of spontaneously generating dislocations 
from defects such as sessile dislocation loops and grain boundaries
\cite{pfcunifieddefects2014}, 
cylindrical sources are input here ``by hand'' primarily for convenience.
They facilitate relatively controlled dislocation nucleation behavior
at stresses below the homogeneous nucleation threshold and from
sources small enough to introduce several within a single simulation cell.
The cylinder acts as a rigid, uniform body similar to a large glass
bead in a colloidal crystal or an incoherent inclusion in an atomic crystal.
When a sufficiently large stress $\sigma_A$ is applied as 
previously discussed,
the cylinder becomes a site for heterogeneous nucleation of 
$\vec{b}=a\langle 100\rangle$ dislocation
pairs or sets of orthogonal pairs
(see Supplemental Material \cite{footnote3} for animation).
Since the stresses are axial, these dislocations nucleate and translate by
climb alone.
We can therefore characterize the creep response
due to purely diffusion-mediated dislocation motion.
In the case shown in Fig.\ \ref{wallpics}, hard wall boundaries are
also employed by slaving $n(\vec{r})$ near the system edges to the
perfect bcc solution \cite{mpfc09}.

If the strain decomposition 
$\epsilon(t) = \epsilon^{\rm el} + \epsilon^{\rm pl}(t)$
is employed once more for the constant $\sigma_A$ condition,
then $\epsilon^{\rm pl}(t)$ can be quantified using the Orowan equation
when the only plastic strain relief mechanism is dislocation motion, 
as noted previously.
If $N$ and $\ell_i$ are both constant
and the average steady-state dislocation velocity can be written
$v_{\rm ss} \sim \sigma_A^q$, then
$\dot{\epsilon}^{\rm pl} \sim \sigma_A^q$.
For pure climb, $q=1$ is generally assumed and has been verified to
hold in PFC simulations \cite{pfcdisloc06}.
Thus we expect $m=q=1$ in constant stress
experiments for which the 
dislocation density $\rho_d=N/V$ is small and constant and the overall stresses
and dislocation velocities are low and linear, 
respectively.
This should approximately be the case between
nucleation of the first and second waves of dislocation pairs
in large $L_y,L_z$ systems.
Analyses and comparisons of observed nucleation rates and pile-up spacings
with those predicted by continuum descriptions are outlined in
Appendices \ref{appendix1} and \ref{appendix2}, respectively. 
The agreement is good in both cases.

\begin{figure}[btp]
 \centering{
 \includegraphics*[width=0.48\textwidth,trim=0 0 0 0]{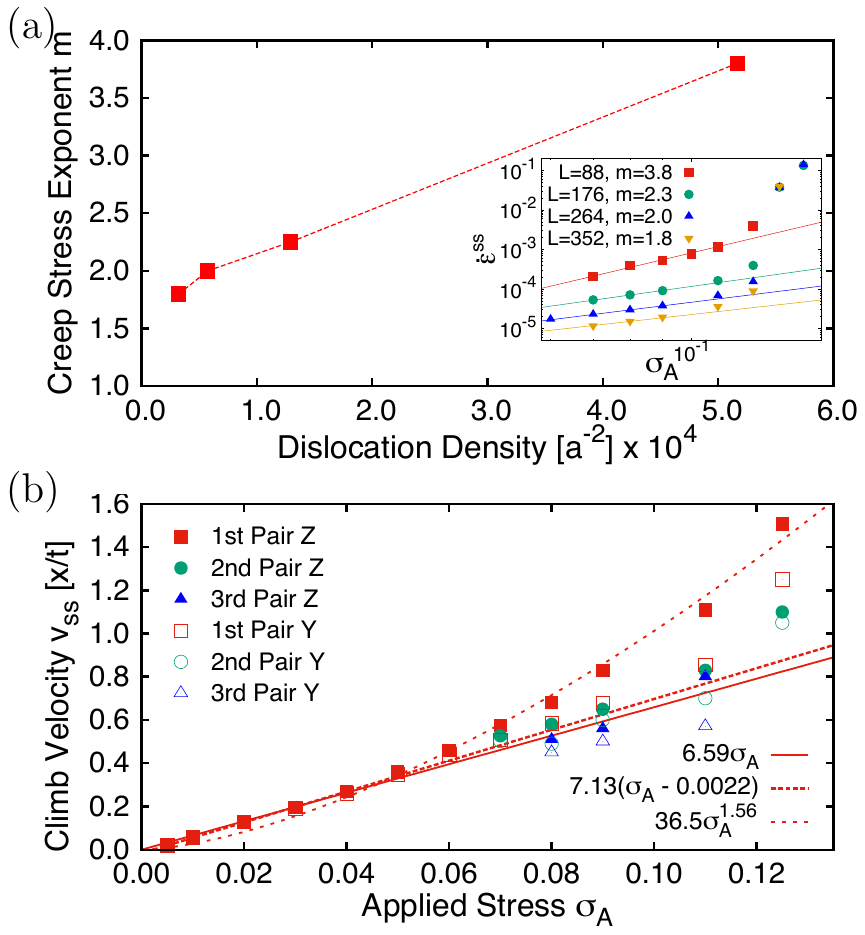}
 }
 \vspace{-.5cm}
\caption[]
{\label{mvsL}
(Color online)
(a) Stress exponent $m$ for simple dislocation creep
vs.\ dislocation density $\rho_d$.
Inset: Steady state strain rate data from which $m$ values were determined
($L=L_y=L_z$).
(b) Steady-state dislocation velocity $v_{\rm ss}$ vs applied stress
$\sigma_A$ for the first 6 dislocation pairs 
in the lowest $\rho_d$ system examined.
The red lines are fits to the first 
pair of nucleated dislocations
that climb along $\pm \vec{z}$ (see Fig.\ \ref{wallpics}(a)).
}
\end{figure}

Figure \ref{mvsL}(a) shows the measured dependence of $m=q$
on system size or
dislocation density $\rho_d$.
The exponent appears roughly to approach 1 as $\rho_d \rightarrow 0$, 
though system size limitations prevent us from accessing very low
$\rho_d$ values.
For easily accessible $\rho_d$, these simple dislocation creep
simulations exhibit $m \gtrsim 2$.
The deviation from $m=1$ appears to have two primary causes.
The first, as already noted, is large $\rho_d$.
For a lattice constant of $0.33$nm
(as assumed in the following),
the densities simulated correspond to 
$\rho_d \simeq 10^{14}/$m$^2 - 10^{16}/$m$^2$, which is near or above
the highest average values typically observed in experiments
($\sim$$10^{12}/$m$^2$).
Our simulations therefore indicate that
collective interactions between 
defects 
at large $\rho_d$ 
can significantly accelerate plastic flow
relative to that expected for isolated dislocations.
A total $\rho_d \lesssim 10^{12}/$m$^2$ should therefore ideally
be maintained in such simulations,  
though 
large \textit{local} densities
$\rho_d \sim 10^{12}/$m$^2$ to $10^{16}/$m$^2$
may have direct physical relevance in some cases.

The other source of $m>1$ is associated with
nonlinear acceleration of $v_{\rm ss}$ at large $\sigma_A$,
as shown in Fig.\ \ref{mvsL}(b).
This acceleration correlates with the onset of significant nonlinear
elasticity and can be avoided for $\sigma_A \lesssim 0.05$, but the existence of
a finite stress barrier $\sigma_N$ for dislocation nucleation
($\sigma_N \simeq 0.035$ in this case)
also limits the explorable range at low $\sigma_A$.

If we follow the analysis of Ref. [\onlinecite{kabirclimbKMC10}] by
applying Taylor's relation directly to our measured climb velocities,
then the predicted creep rates would follow
$\dot{\epsilon} \sim \rho_d b v_{ss} \sim \sigma^{2+[1.8,3.8]} 
\sim \sigma^{[3.8,5.8]}$ for the values of $\rho_d$ examined.
These values are comparable to those obtained from the 
atomistically-informed
kMC simulations of Ref. [\onlinecite{kabirclimbKMC10}], $m \simeq 5-5.5$,
which were conducted at dislocation densities near the high end of our 
$\rho_d$ range.
The underlying causes of these large exponents
were identified in their study
as nonlinear vacancy-dislocation core interactions
and a non-negligible dependence of vacancy concentration on stress.
Such mechanisms are not explicitly considered in PFC descriptions,
but nonlinearities near dislocation cores are naturally captured by
$n(\vec{r})$, the local peaks of which decrease in amplitude and increase
in width as the core is approached.
These variations signify local changes in vacancy concentration and migration
barriers, which appear to produce qualitatively similar 
nonlinearities in climb rates with increasing $\rho_d$ 
to those observed in the atomistic simulations of 
Ref. [\onlinecite{kabirclimbKMC10}].

For sufficiently low $\rho_d$ ($\lesssim 10^{13}/m^2$), our PFC description
appears to converge toward the standard
description of purely climb-mediated creep in which $q=1$ and thus $m=3$.
The noted nonlinearities caused by interacting dislocation strain
fields apparently become negligible at and below typical physically relevant
dislocation densities.
This finding is in accord with Clouet's analysis \cite{clouetclimbPRB11}
of the results of Ref. [\onlinecite{kabirclimbKMC10}], wherein $m \simeq 3$
is obtained for experimentally relevant conditions.
It therefore appears that much
larger, less controlled system configurations in which dislocations
may more realistically nucleate, climb, glide, annihilate, etc.
must be examined to obtain a more meaningful description of power law creep.
The required length scales are not yet accessible with PFC simulations,
but we initiate steps in this direction in the following section.
We note nonetheless that the overall consistency of our results 
with those of atomistic 
and mesoscale descriptions indicates that both the atomistic 
mechanisms and long-range interactions that underlie dislocation climb are
qualitatively captured by PFC models.

\subsection{Power law creep}
\label{powerlawcreep}

\begin{figure}[btp]
 \centering{
 \includegraphics*[width=0.48\textwidth,trim=0 0 0 0]{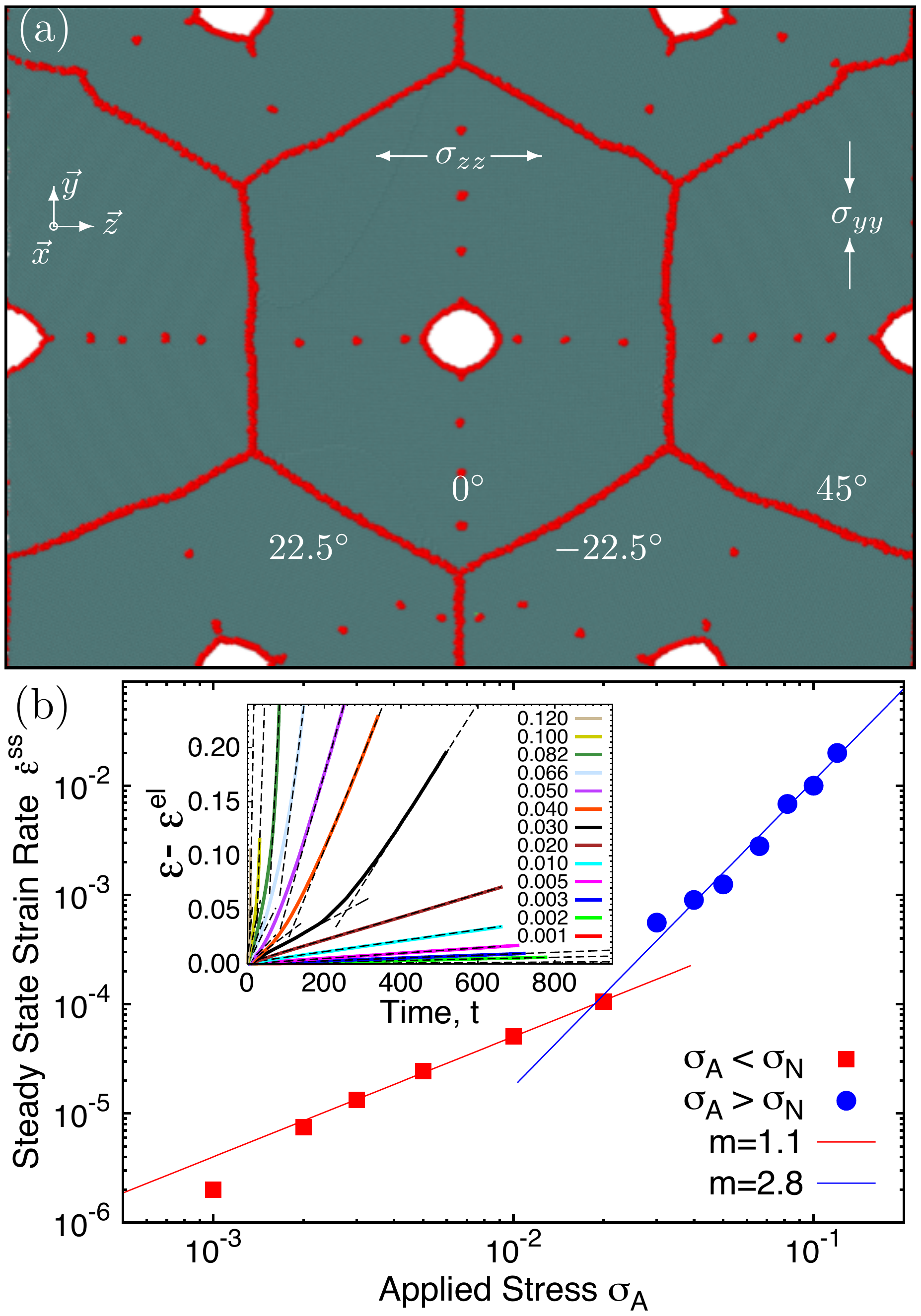}
 }
 \vspace{-.5cm}
\caption[]
{\label{hexincl}
(Color online)
Combined dislocation and $22.5^{\circ}$/$45^{\circ}$ grain boundary creep
at $d=152.5a$ with 1 central 
dislocation 
source per grain.
(a) System representation at $\sigma_A=0.04$ and
$t=160$, with colors as in Fig.\ \ref{wallpics}.
(b) Steady state strain rate data and stress exponents $m$. 
Inset: The corresponding creep curves at various $\sigma_A$ 
(color legend).
}
\end{figure}

We next examine somewhat less idealized microstructures
to determine whether 
more varied modes of dislocation creation, interaction, and annihilation
can self-consistently generate power law creep behavior
within atomistic PFC descriptions.
As a first case,
dislocation sources are symmetrically 
introduced into the hexagonal grain structures
of Section \ref{sec:hexcreep},
as shown in Fig.\ \ref{hexincl}(a).
This configuration produces a crossover from
$m \simeq 1$ diffusional creep for $\sigma_A < \sigma_N$ to 
$m \simeq 2.8$ for $\sigma_A > \sigma_N$, as shown in 
Fig.\ \ref{hexincl}(b)
(see Supplemental Material \cite{footnote3} for animation).
Below $\sigma_N$, dislocations do not nucleate and the response is unchanged
from that outlined in Section \ref{sec:hexcreep} for diffusional creep and 
stress-assisted grain boundary migration. 
Above $\sigma_N$, dislocations nucleate from the sources and the response 
therefore transitions toward a dislocation mediated power-law-type regime.
This change in strain relief mechanisms above $\sigma_N$ is clearly visible,
with a crossover to larger but still nearly linear steady-state strain rates 
after dislocations become active.
The mobile dislocation density becomes large during this regime,
on the order of $\rho_d \simeq 10^{15}/$m$^2$,
and increases roughly as $\rho_d \sim \sigma_A$.
The observed exponent $m \simeq 2.8$ therefore seems to
result primarily from the combined effects of this
$\rho_d \sim \sigma_A$ dependence
and the large $\rho_d$ and nonlinear $v_{\rm ss}$ effects,
as illustrated in Fig.\ \ref{mvsL}.
Glide is active in these systems, but does not appear to contribute
significantly to the plastic flow in this idealized system geometry.

A second system configuration containing
16 randomly positioned sources with $L_y=L_z=352a$
and 4 hard wall boundaries
is shown in Fig.\ \ref{randwall}
(see Supplemental Material \cite{footnote3} for animation).
Such systems nucleate waves of dislocations when $\sigma_A > \sigma_N$,
and between early waves generally reach temporary steady-states in terms of
$\rho_d^{\rm ss}$ and $\dot{\epsilon}^{\rm ss}$.
Between the first and second waves of dislocation nucleation, for example,
when $\rho_d \simeq 3\times 10^{-15}/m^2$ (at all stresses)
we obtain $m \simeq 4.2$ (Fig.\ \ref{randwall}(b)).
This value is larger than that expected from large $\rho_d$
and nonlinear $v_{\rm ss}$ effects, which as quantified in
Fig.\ \ref{mvsL} would 
be expected to 
produce $m \simeq 3.0 \pm 0.4$.
Similar simulations with 2 rather than 4 hard wall boundaries
produce $m \simeq 4.0$. 
The excess $m$ in these simulations may be due to the contribution of 
dislocation glide.

\begin{figure}[btp]
 \centering{
  \includegraphics*[width=0.48\textwidth,trim=0 0 0 0]{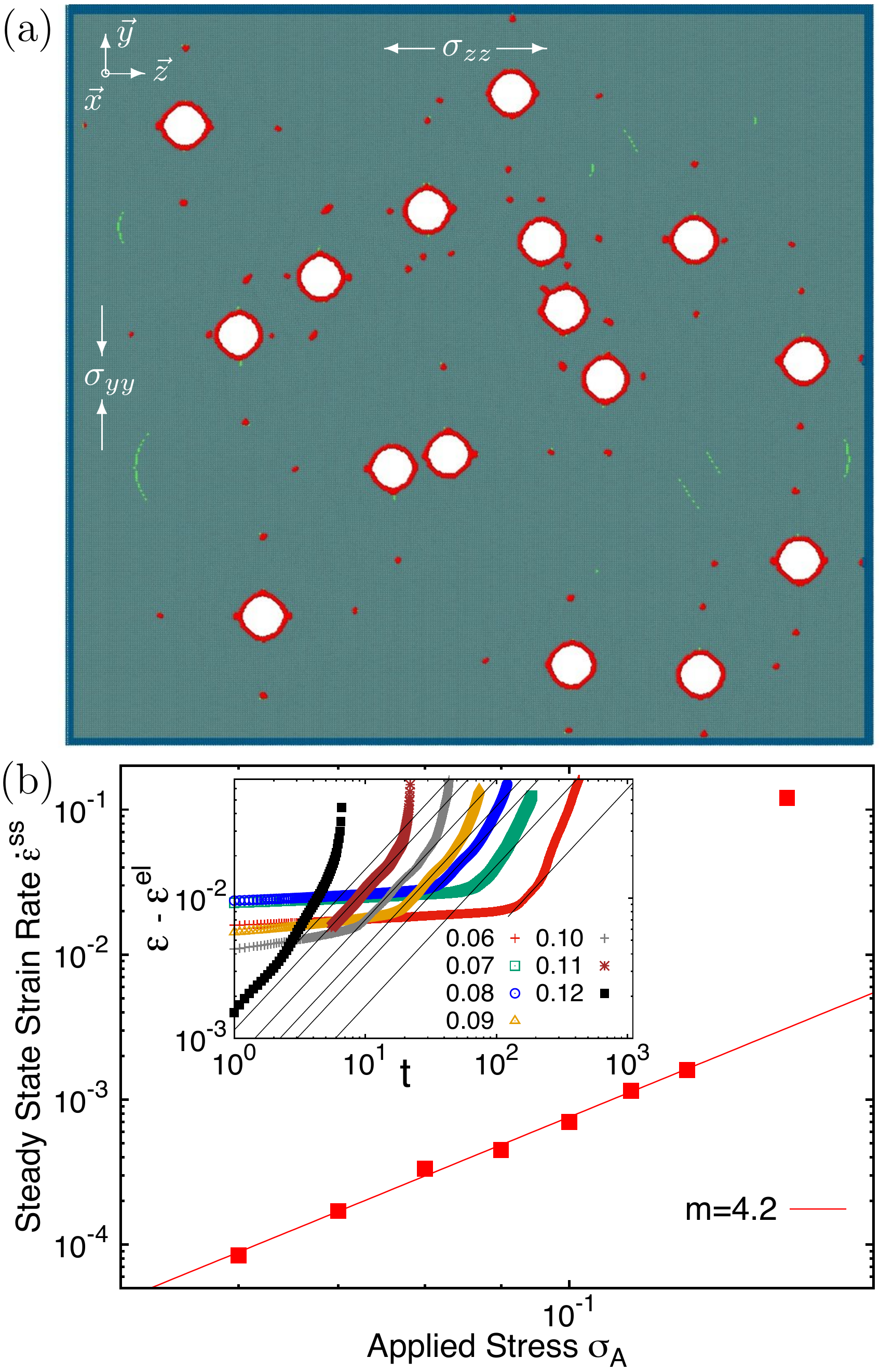}
 }
 \vspace{-.5cm}
\caption[]
{\label{randwall}
(Color online)
Dislocation creep at $L_y=L_z=352a$
with 16 randomly positioned sources
and 4 hard wall boundaries.
(a) System representation at $\sigma_A=0.08$ and $t=65$,
with colors as in Fig.\ \ref{wallpics}.
(b) $\dot{\epsilon}^{\rm ss}$ data from which $m$ was determined.
Inset: Creep curves 
at various $\sigma_A$
for the configuration shown in (a).
}
\end{figure}

If we extrapolate these results to lower $\rho_d$ where $q=1$ is expected
to be realized (decrement measured $m$ by $\sim$2), 
and further assume that Taylor's relation holds (increment measured $m$ by 2), 
a power law creep stress exponent of $m \simeq 4.0-4.2$ is implied.
This value approaches but is smaller than 
typical experimental results for pure metals,
$m \gtrsim 4.5$, and is approximately consistent with 
previous mesoscopic descriptions
that consider both climb and glide ($m=4.5$) \cite{weertman75}.
These findings, though not obtained directly at the system sizes and dislocation
densities ulitimately sought, do indicate that physically meaningful
behavior is being captured.

As noted previously, 
the power law exponents
obtained here at large $\rho_d$ may also 
have some limited direct physical relevance
based on the known spatial heterogeneity of $\rho_d$ in many systems.
Even though \textit{average} values of 
$\rho_d \gtrsim 10^{-5}/a^2 \simeq 10^{14}/$m$^2$
simulated here are not typically encountered in real materials,
significant spatial heterogeneity in $\rho_d(\vec{r})$ is often observed.
Dislocation clustering at cell walls, tangles, obstacles,
and even low-to-medium angle grain boundaries, for example,
have been estimated \cite{disloccells83}
to produce local values of $\rho_d \sim 10^{16}/$m$^2$,
as large as any considered here.
Though these heterogeneous domains of very large $\rho_d$ are separated
by domains of much lower $\rho_d$ 
($\sim 10^{13}/$m$^2$ in Ref.\  [\onlinecite{disloccells83}]), 
they may nonetheless locally induce strong deviations from
the $q=1$ climb kinetics commonly assumed for individual dislocations,
based on the simulations of Section \ref{subsec:disloctheory}.
Apparent power law creep behavior may therefore be observed 
in such systems, even in the absence
of any particularly complex defect reactions or mesoscale patterning.

\section{Conclusions}
\label{sec:conclusions}

We have studied the nonequilibrium kinetics of creep deformation
and diffusional defect migration
in atomistic crystalline and nanopolycrystalline systems using
phase field crystal modeling.
A method for conducting constant stress PFC simulations,
the first such method to our knowledge,
was developed and utilized throughout this study.
The characteristic stress and grain size exponents
quantified for symmetric nanopolycrystals,
$m \simeq 1.02$ and $p \simeq 1.98$, respectively, 
closely match those expected for idealized diffusional
Nabarro-Herring creep, 
$m=1$ and $p=2$.
We find that a significant portion of the plastic flow in these systems is
associated with non-affine grain boundary motion, suggesting that 
concurrent stress-assisted diffusive boundary migration 
does not necessarily alter diffusional creep exponents 
in the regime of large boundary mobilities.
Exponents consistent with
diffusional Coble creep ($m=1$, $p=3$)
were observed only in the presence of stochastic thermal noise 
amplitudes $M$ large enough to induce grain boundary melting or premelting.
This suggests that a re-entrant transition 
to Coble creep may occur in weakly stressed systems that exhibit significant
premelting near $T_m$.

In simulations of dislocation-mediated plastic flow
(Section \ref{subsec:disloctheory}),
nucleation rates and pile-up spacings were shown
to be well described by 
predictions of continuum elastic dislocation theory,
Eqs.\ (\ref{nucleq}) and (\ref{pileupeq}), 
respectively.
At numerically accessible dislocation densities, strong nonlinear interactions
were found to produce `inherent' dislocation climb exponents 
$1.8 \lesssim q \lesssim 3.8$
and extrapolated creep stress exponents
$3.8 \lesssim m \lesssim 5.8$
for $\rho_d \sim 10^{14}/a^2$ to $10^{16}/a^2$.
These results are in general agreement with previous 
atomistic \cite{kabirclimbKMC10}
and mesoscale studies \cite{clouetclimbPRB11}, and further highlight the need
to access lower dislocation densities in atomistic simulations.
Extrapolated climb rates appear to approach $q \simeq 1$
near typical experimental dislocation densities
($\rho_d \sim 10^{12}/a^2$),
indicating that
PFC models will reproduce climb-mediated natural creep behavior with
$m=3$ when Taylor's relation holds.

In systems with slightly less idealized grain and dislocation source stuctures
(Section \ref{powerlawcreep}),
stress exponents $m \simeq 2.8-4.2$ were directly measured, but
these values cannot be straightforwardly compared to experimental results due
to the known influence of large $\rho_d$ effects.
Anomalously large stress exponents associated with the kinetic climb rate
nonlinearities observed at large $\rho_d$
may nonetheless have relevance to power law creep in systems that exhibit
comparable dislocation densities locally within dislocation tangles and 
cell walls, for example.
Extrapolation of the simulation results 
of Section \ref{powerlawcreep} to physically 
relevant dislocation densities 
leads to predicted exponents of $m \simeq 4.0-4.2$, nearly consistent with 
mesoscale predictions
for systems in which both climb and glide are active ($m=4.5$), but still
somewhat smaller than typical experimentally measured values in pure metals
($m \gtrsim 4.5$).

The results of this work indicate that PFC-type approaches are capable of
capturing the essential physics of nonconservative defect evolution and
dislocation-mediated creep at atomistic length scales and diffusive time scales.
The system sizes required to describe typical experimentally-relevant
dislocation densities, though not yet accessible, may be realizable
within the next decade or potentially sooner with further development of
computationally efficient, coarse grained complex amplitude expansions 
of PFC models
\cite{pfcadmesh07,pfcbinaryamp10,pfcRGkarma10,xpfcamp13}.
Potential points of shorter-term study include investigation of
fully 3D systems with more realistic microstructures
and a distribution of dislocation source activation stresses,
as well as development of constant stress deformation methods
that impose fewer constraints on the shape of the
polycrystalline domain.

\vspace*{-0.5cm}
\begin{acknowledgments}
This work has been supported by the Natural Science and Engineering 
Research Council of Canada (NSERC), and access
to supercomputing resources has been provided by CLUMEQ/Compute Canada.
\end{acknowledgments}

\appendix

\section{Dislocation Nucleation Rates}
\label{appendix1}

As dislocation pairs sequentially nucleate
in the simulations of Section \ref{subsec:disloctheory},
the dependence of $\rho_d$ on time and $\sigma_A$ can be
understood as follows.
Upon nucleation of the first dislocation pair at $t_1$, the local strain around
the source is reduced by $\epsilon_{zz}=b/L_z$.
For $\sigma_A \simeq \sigma_N$,
this amount of strain must be reapplied to nucleate another pair,
and the time required to do so, $t_2-t_1$, is equal to the time it takes the
climbing dislocation pair to plastically relieve this same amount of strain.
For a constant steady-state climb velocity,
\be
t_2-t_1 = \frac{b}{L_z \dot{\epsilon}^{\rm pl}_{zz}} =
\frac{L_y}{2 v_{\rm ss}}.
\ee
At $t_2$ there are 4 mobile dislocations, so the time required to
nucleate the third pair is halved, and so on.
In general,
\be
t_i-t_{i-1} = \frac{b}{L_z \dot{\epsilon}^{\rm pl}_{zz}} =
\frac{L_y}{2(i-1) v_{\rm ss}}
\ee
and the time required to generate $i$ pairs of dislocations is
\begin{eqnarray}
t_i &=& t_1 + \frac{L_y}{2v_{\rm ss}}
+ \frac{L_y}{4v_{\rm ss}}
+ \cdots 
+ \frac{L_y}{2(i-2) v_{\rm ss}}
+ \frac{L_y}{2(i-1) v_{\rm ss}}
\nonumber\\
&=&t_1+
\frac{L_y}{2 v_{\rm ss}}\left[
1 + \frac{1}{2}
+ \cdots
+\frac{1}{i-2} 
+\frac{1}{i-1} 
\right]
\nonumber\\
&=&t_1+\frac{L_y}{2v_{\rm ss}}\sum_{n=1}^{i-1} \frac{1}{n}
=t_1+\frac{L_y}{2c\sigma_A}\sum_{n=1}^{i-1} \frac{1}{n}
\label{nucleq}
\end{eqnarray}
where $c\simeq6.59$ is the constant prefactor given in Fig.\ \ref{mvsL}(b).
Equation (\ref{nucleq}) is a harmonic series, 
which grows faster than logarithmically,
implying that $\rho_d$ diverges faster than exponentially
when annihilation and/or immobilization do/does not occur
\cite{footnote4}.
This would be the case, for example, in an infinitely large system.
Since our simulations are finite and also employ hard wall boundaries, 
dislocations become effectively immobile when they reach a wall.
Nucleation of a new pair tends to
occur roughly when the previous pair reaches the hard wall boundary,
such that the number of mobile dislocations $N$ remains nearly constant
with time. 
For this type of fixed $N$ scenario,
Eq.\ (\ref{nucleq}) predicts that periodic waves of dislocation nucleation 
will occur every $L_y/(2v_{\rm ss})$
such that the total number of mobile plus immobile dislocations 
increases linearly with time at a
rate proportional to $\sigma_A/L_y$.
These relations are indeed found to describe the first few
waves of nucleation quite well
for all system sizes studied with hard wall boundaries.

\section{Dislocation Pile-Up Spacings}
\label{appendix2}

\begin{figure}[H]
 \vspace{.5cm}
 \centering{
 \includegraphics*[width=0.48\textwidth,trim=0 0 0 0]{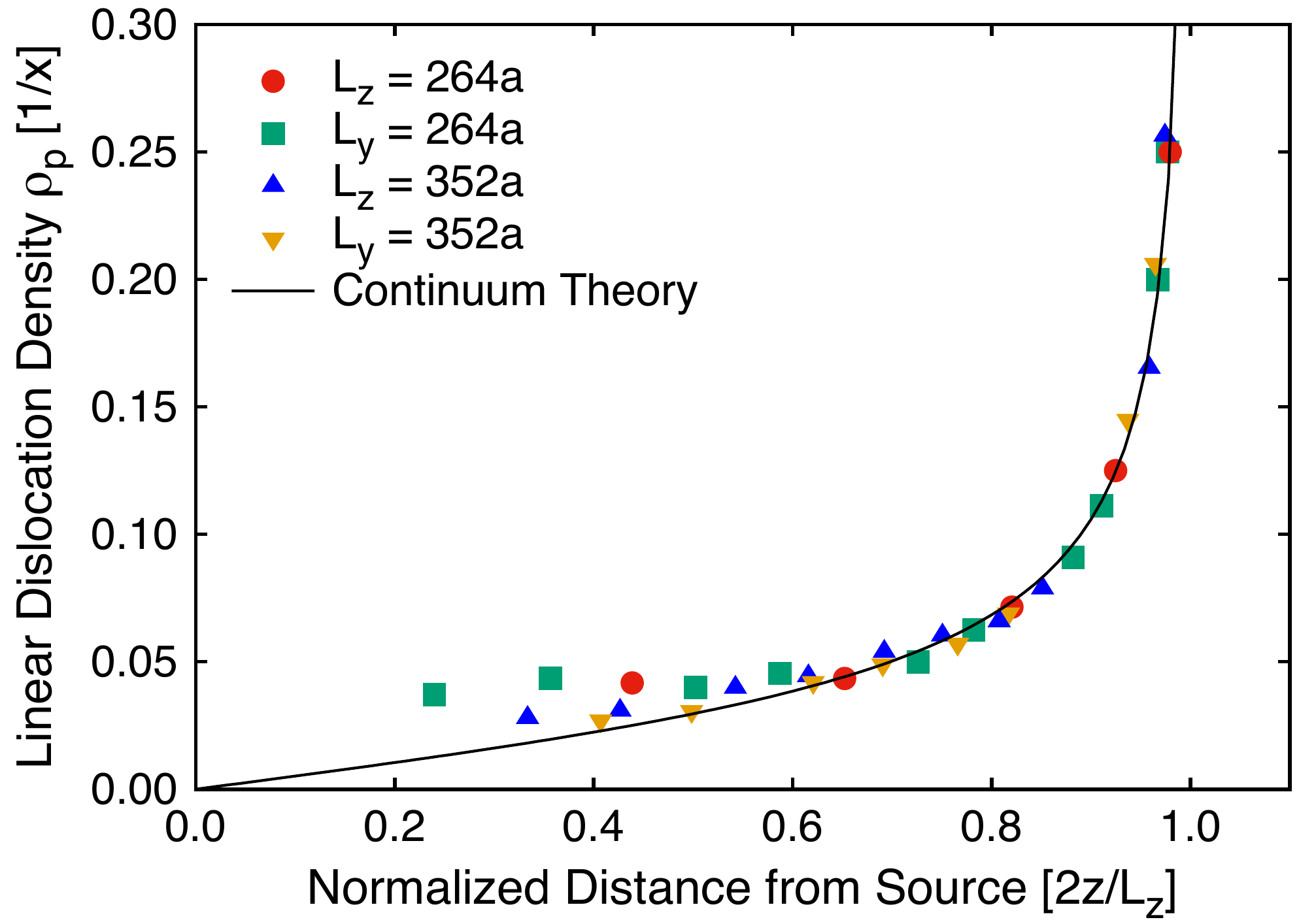}
 }
 \vspace{-.5cm}
\caption[]
{\label{pileupspacing}
(Color online)
Comparison of dislocation pile-up spacing with predictions of continuum theory.
Simulations are at $\sigma_A=0.045$ and various $L_y=L_z$.
}
\end{figure}

Figure \ref{pileupspacing} shows the linear dislocation density
$\rho_p(z) = \int_{z-dz}^{z+dz} N(z')dz'/(2dz)$ 
as a function
of distance from the source for pile-ups such as that seen in
Fig.\ \ref{wallpics}(c).
$\rho_p(z)$ is defined such that
the number of dislocations between $z-dz$ and $z+dz$
is $2 dz \rho_p(z)$.
The solid line is a best fit to the prediction of continuum
elasticity theory \cite{hirthlothe} for a double 1D pile-up along $z$, 
\be
\rho_p(z) = \frac{2(1-\nu)\sigma_A}{\mu b}\frac{z}{\sqrt{(L_z/2)^2-z^2}}
\label{pileupeq}
\ee
where $z$ is the distance from the center of the pile-up.
This equation 
with $\nu=1/2$ (as imposed by the fixed area, constant $\sigma_A$ algorithm), 
$\sigma_A=0.045$, $\mu=1$, and $b=\sqrt{2/3}$
accurately describes the simulation results for $z$ sufficiently 
far from the source.
This further confirms that the long range elastic interactions between
PFC dislocations agree with continuum elastic descriptions,
such that non-trivial multi-dislocation, multi-obstacle interactions 
are captured as well.

\begin{thebibliography}{44}
\bibitem[{\citenamefont{footnote1}(2015)}]{footnote1}\citenamefont{
For simplicity, a $k=0$ mode is not considered, although we do not expect 
its effect to change any of the physics reported herein. 
}
\bibitem[{\citenamefont{footnote2}(2015)}]{footnote2}\citenamefont{
If a constant area or volume constraint is not used, then
one must continuously adjust the total system number density $n_0$
in inverse proportion to the volume change, such that
the total number of particles remains fixed.
It is not clear whether this can be done consistently with a PFC description,
as an exact relation between $n_0$ and number of particles is
in general not known.
If $n_0$ is not adjusted with changes in volume, 
the volume can change unphysically without bound or penalty
as local density peaks or lattice sites are simply created or removed
to compensate for the volume change.
}
\end{thebibliography}

\end{document}